\documentstyle[aps,epsfig]{revtex}
\begin{document}
\draft
\title{Accurate Results from Perturbation Theory for Strongly Frustrated
       $S=\frac{1}{2}$ Heisenberg Spin Clusters}

\author{N.P. Konstantinidis and D. Coffey}

\address{Department of Physics, State University of New York at Buffalo,
Amherst, NY 14260}

\date{\today}
\maketitle

\begin{abstract}

We investigate the use of perturbation theory in finite sized frustrated spin
systems by calculating the effect of quantum fluctuations on coherent states
derived from the classical ground state. We first calculate the ground and
first excited state wavefunctions as a function of applied field for a 12-site
system and compare with the results of exact diagonalization. We then apply the
technique to a 20-site system with the same three fold site coordination as the
12-site system. Frustration results in asymptotically convergent series for
both systems which are summed with Pad\'e approximants.

We find that at zero magnetic field the different connectivity of the two
systems leads to a triplet first excited state in the 12-site system and a
singlet first excited state in the 20-site system, while the ground state is a
singlet for both. We also show how the analytic structure of the Pad\'e
approximants at $|\lambda| \simeq 1$ evolves in the complex $\lambda$ plane at
the values of the applied field where the ground state switches between spin
sectors and how this is connected with the non-trivial dependence of the
$<S^{z}>$ number on the strength of quantum fluctuations. We discuss the origin
of this difference in the energy spectra and in the analytic structures. We
also characterize the ground and first excited states according to the values
of the various spin correlation functions.
\end{abstract}


\section{Introduction}

The antiferromagnetic spin Heisenberg model has been the object of intense
investigation through the years. Recently it has attracted enormous interest in
the study of strongly correlated electron systems, which include the oxide
superconductors \cite{Manousakis} and low dimensional spin systems. In the
limit where the on-site Coulomb repulsion is very strong, it is equivalent to
the one band Hubbard model for half filling. The inclusion of competing
interactions in it has led to novel quantum phases, making it appropriate for
the study of quantum criticality \cite{Sachdev}.

The solution of the model was calculated by Bethe in one dimension for nearest
neighbor interactions \cite{Bethe}, but a solution in analytic form is lacking
for two or three dimensions, except for special cases \cite{Caspers}.
Approximation and numerical techniques that have been used include
diagonalization of small clusters \cite{Bernu,Lhuillier}, Monte Carlo
techniques \cite{Johnston}, cluster expansions \cite{Singh}, spin wave
expansions \cite{Mattis,Zhong,Trumper} and the density matrix renormalization
group \cite{White}. The methods that consider the full Hilbert space of the
problem are limited by the size of the system, since the number of states is
exponentially dependent on it. In Monte Carlo calculations the sign problem
leads to loss of statistical accuracy, especially for frustrated systems
\cite{Johnston}.

An alternative approach to these techniques is the direct application of
perturbation theory in which corrections to the classical treatment are
calculated order by order in the strength of residual interactions, the effect
of fluctuations left out in the mean field approximation. Although this
approach can also be limited in the size of the systems which can be
investigated due to the dimensionality of the Hilbert space, it is very
different in that it provides analytic information on the effects of
corrections to classical approximations and so complements the other approaches
mentioned above. We use this approach to calculate the ground state
wavefunction of two systems which have qualitatively different classical ground
states, as we will explain below. We demonstrate that the differences survive
in the exact spin $\frac{1}{2}$ ground states through the difference in the
analytic structures of the Pad\'e approximants derived from the perturbation
expansions.

%The model has been investigated for various
%lattices, such as the square, the triangular and the Kagom\'{e} lattices. The
%square lattice has no frustration, while for the other two competing
%interactions lead to frustration, complicating the problem. The properties of
%interest for these lattices are the nature of the ground state and the
%excitation spectrum, the existence of magnetic order and possible phase
%transitions driven by the strength of coupling constants. The approximation of
%nearest neighbor interactions is usually employed.

We consider the Heisenberg model on closed two dimensional spin $\frac{1}{2}$
systems which have three nearest neighbors. The antiferromagnetic interaction
between the spins leads to frustration at the classical level.
%Frustrated systems have been found to possess non-trivial low temperature
%properties \cite{Ramirez}.
One such system is the 60 site cluster whose
relative positions are the same as those of carbon atoms in $C_{60}$ (from now
on we will refer to n site systems with the spatial symmetry of the fullerenes
as $C_{n}$). The 60 site system consists of 20 hexagons and 12 pentagons.
Assuming a tight binding model for its electronic properties where there is one
orbital per site and a strong Coulomb repulsion for double occupancy, the
Heisenberg model on it gives an effective low energy description at half
filling. This is a first approximation to the problem, since $C_{60}$ is
estimated to be in the intermediate and not in the strong coupling $U$ regime
\cite{Coffey}. The hopping matrix elements between sites on the same pentagon
can be taken to be different from the ones between sites on adjacent pentagons.
This leads to two positive exchange constants $J_{1}$ and $J_{2}$ and the
Hamiltonian is:
\begin{equation}
 H = J_{1} \sum_{<i,j>}^{p.} \ \vec{S}_{i} \cdot \vec{S}_{j} +J_{2} \sum_{<i,j>}^{n.p.} \ \vec{S}_{i} \cdot \vec{S}_{j} \hspace{5pt}.
\end{equation}
$J_{1}$ refers to bonds between the same pentagon and $J_{2}$ to non-pentagon
bonds, while $<>$ stands for nearest neighbor interactions.

%The classical ground state of the Heisenberg Hamiltonian has been determined on
%this and analogous systems and has been found to have a strong dependence on
%connectivity in its response to a magnetic field \cite{Coffey}. In particular
%in $C_{20}$ and $C_{60}$ it has discontinuities in the magnetization $M$, as a
%function of the magnetic field, whereas in $C_{70}$ and $C_{84}$ it has
%discontinuities in the magnetic susceptibility
%$\frac{\partial{M}}{\partial{H}}$. The magnetization of the quantum mechanical
%ground state jumps when there is a transition at critical values of the
%magnetic field from one $S^{z}$ sector to the next one. We examine if there is
%any correlation between these values and the ones where the classical jump in
%the magnetization (or the susceptibility) takes place.

The classical ground state of the Heisenberg Hamiltonian has been determined on
this and analogous $C_{n}$ systems \cite{Coffey}. The magnetic properties of
these ground states show an unexpected dependence on $n$. In some the
magnetization is discontinuous in an applied field whereas in others it is the
susceptibility which is discontinuous. It is of interest to determine whether
this dependence is an artifact of the classical approximation or whether it is
present in $S=\frac{1}{2}$ solutions. We use the classical $C_{n}$ ground
states to define an Ising Hamiltonian where the quantization axis at each site
is determined by the direction of the spin vector in the classical ground
state. Coherent states are defined along these axes and constitute a mean field
approximation for the quantum ground state. In this way, each site has a local
axis associated with it, in contrast to the basis where all spins are defined
in the same coordinate system in spin space. From now on we call the former
'local basis', and the latter 'global basis'. The quantum fluctuations are then
built around the classical directions. They are the terms added to the original
mean field approximation, and they are multiplied with a parameter $\lambda$
which varies from 0 to 1. When $\lambda=1$, the full isotropic Heisenberg
Hamiltonian is recovered. Thus the solution is generated as a series expansion
in the perturbation parameter $\lambda$ with application of perturbation
theory, and the expansions of the ground state energy and wavefunction in the
local axes basis are known. With this approach we can study the evolution of
the system away from the classical ground state and towards the full quantum
limit $\lambda=1$.

However, the Hilbert space for the 60 site system is huge, consisting of
$2^{60} = 1.15 \times 10^{18}$ states, so a perturbation treatment in the whole
Hilbert space can only give a few orders. This is because the number of states
involved in the calculation rapidly increases as the order and frustration
increase \cite{Modine}. In addition, since the spin axes are directed along the
classical solution's directions, the total spin in the global basis is not a
good quantum number of the Hamiltonian (when $\lambda<1$), so a reduction of
the number of states by focusing on a particular $S^{z}$ value is not possible.
Another way of gaining insight into the problem is to consider similar smaller
systems, belonging to the same family. In all these systems there are
$n_{h} = \frac{n}{2} -10$ number of hexagons and 12 pentagons. The smallest
member of the group is the 20 site system. Again, we consider $S=\frac{1}{2}$
spins sitting at their vertices.

%${60 \choose 30} = 1.18 \times 10^{17}$

The frustration of the Hamiltonians studied leaves its signature in the various
series expansions generated by perturbation theory, producing functions with
non-analytic structure in the complex $\lambda$ plane. The presence of branch
cuts limits absolute convergence within a circle centered at the origin with a
radius of convergence smaller than 1 and the series are only asymptotically
convergent in the full isotropic limit where $\lambda=1$. Therefore, we employ
analytic continuation with the use of Pad\'e approximants. The structure of the
functions in the $\lambda$ plane depends strongly on the form of the perturbing
Hamiltonian. We investigate the signature of the changes in the complex plane
structure of the related functions as the perturbing Hamiltonian is varied. The
generalization to complex variables has been proven useful in the study of
phase transitions in the two dimensional Ising model in temperature $T$ in a
complex magnetic field \cite{Lee}. The systems studied here are closed and do
not possess a thermodynamic limit. However, knowledge of the structure in the
complex coupling constant plane provides information about the functions
studied \cite{Matveev}. It can also provide information for the evolution of
instabilities. Here the only possible transitions are between $<S^{z}>$
sectors with increasing magnetic field, where $<S^{z}>$ is the expectation
value of the z component of the spin in the global basis.

The classical ground state at zero magnetic field is doubly degenerate, since a
flipping of all the spins does not change the energy. Consequently degenerate
perturbation theory has to be applied \cite{Gelfand}. This is done via a
similarity transformation, and an effective $2\times2$ matrix is generated.
This matrix provides information on the ground and first excited state,
including the evolution of $<S^{z}>$ as a magnetic field is varied. The
structure of this matrix in the complex $\lambda$ plane can be correlated with
the transitions of the $<S^{z}>$ number between different sectors as a function
of the magnetic field.

The plan of this paper is as follows: in section II the method for the solution
of the problem using perturbation theory is described. In section III this is
applied to a twelve site system, $C_{12}$, where the results are compared with
exact diagonalization and found to be in complete agreement. We also discuss
how the analytic structure of perturbation theory is reflected in the magnetic
properties. In section IV perturbation theory is applied to $C_{20}$ and the
results are tested by recovering expectation values for $\vec{S}^{2}$ and
${S}^{z}$ very close to integer values. Here we contrast the magnetic field
dependence of the analytic structure with the results found in the classical
approximation. For both $C_{12}$ and $C_{20}$ ground and first excited state
wavefunctions are calculated in applied magnetic fields and the dependence of
their magnetic properties on the strength of quantum fluctuations is
determined. In the case of $C_{20}$, the ground and excited states at zero
magnetic field are singlets. This result is in agreement with similar ones for
strongly frustrated magnetic systems such as the Kagom\'e lattice and a one
dimensional analogue of the pyrochlore lattice \cite{Mambrini}. The
non-magnetic nature of the excitation is attributed to the frustration and the
connectivity of the system.

\section {METHOD}

The starting point in the calculation is the classical ground state. The
Hilbert space is spanned by spin $\frac{1}{2}$ spinors determined by the
classical solution. Local $z_{i}$ axes are defined along the classical spin
directions, and spin states are defined at each site such that the expectation
value of the component of the spin along the axis equals its classical value:
\begin{equation}
\langle {\alpha_{i} \atop \beta_{i}} \vert \vec{\sigma} \vert {\alpha_{i} \atop
\beta_{i}} \rangle = \vec{S_{i}} \hspace{5pt},
\end{equation}
where $\alpha_{i}$, $\beta_{i}$ are spinor coefficients and
$\arrowvert \vec{S_{i}} \arrowvert=1$. The coherent states \cite{Klauder} are
products of spin states along these axes:
\begin{equation}
\vert\Psi_{S}\rangle = \prod_{i=1}^{N} \vert S_{i}\rangle = \prod_{i=1}^{N}
\vert {\alpha_{i} \atop \beta_{i}} \rangle \hspace{5pt},
\end{equation}
where N is the number of spins in the system. Here we limit ourselves to these
$2^{N}$ states out of the overcomplete basis of the coherent states. These are
eigenstates of the unperturbed Hamiltonian $H_{0}$ and constitute an
orthonormal basis. The classical ground states are the ones where all the spins
are either 'up' or 'down' so that $H_{0}$ has the form of an Ising Hamiltonian
with respect to the local quantization axes defined by the classical result.
The fluctuations around the local $z_{i}$ axes are raising and lowering spin
operators, defined along the local $x_{i}$ and $y_{i}$ axes. These are included
in the perturbing part $H_{1}$ which we scale with a parameter $\lambda$. Each
$x_{i}$ axis is defined in the azimuthal plane of the corresponding $z_{i}$
axis and perpendicular to it, and then the $y_{i}$ axis is constructed with the
right hand rule. Due to the absence of a global z axis, the Hamiltonian takes a
complicated form. The solution is generated in perturbation theory as a power
series in $\lambda$. Therefore the Hamiltonian has the form:
\begin{equation}
 H = H_{0}+\lambda H_{1} \hspace{5pt},
\end{equation}
where $H_{0}$ is the classical part and $H_{1}$ the quantum fluctuations.

There are two well known issues which complicate the application of
perturbation theory. The first is the double degeneracy of the coherent states
generated from the classical ground state in the absence of an applied magnetic
field. This requires the use of degenerate perturbation theory. To overcome
this problem, an effective Hamiltonian is constructed for the degenerate ground
states, via a similarity transformation \cite{Gelfand}. The Hamiltonian is
transformed to a block diagonal form and its elements, as well as the
coefficients of the Hilbert space states, which contribute to the perturbed
wavefunctions, are expanded as power series in $\lambda$. Then recurrence
relations can be written down for the effective Hamiltonian and the
wavefunctions:
\begin{equation}
 H_{k}^{eff}(m,\ell)=<m|H_{1}|\Psi_{k-1}^{(\ell)}> \hspace{5pt},
\end{equation}
\begin{equation}
 <n|\Psi_{k}^{(\ell)}> = \frac{1}{E_{0}-E{n}} (<n|H_{1}|\Psi_{k-1}^{(\ell)}>-
 \sum_{k^{\prime}=1}^{k-1} \sum_{\ell^{\prime}=1}^{L} H_{k-k^{\prime}}^{eff}
 (\ell^{\prime},\ell) <n|\Psi_{k^{\prime}}^{(\ell^{\prime})}>) \hspace{5pt}.
\end{equation}
Here k is the order of perturbation, $|m>$ a degenerate ground state of
$H_{0}$, $|n>$ a state in the Hilbert space different from $|m>$, $E_{n}$ its
energy at the classical level and $E_{0}$ the classical ground state energy. L
is the dimensionality of the degenerate subspace (here $L=2$) and $\ell$ runs
from 1 to L. The result of this calculation is an $L \times L$ (here
$2 \times 2$) matrix whose eigenenergies are the ground state and the first
excited state. This method can be applied for zero or non-zero magnetic field,
where all the magnetic field terms in the Hamiltonian are included in the
perturbation. If the magnetic field is included in $H_{0}$ non-degenerate
perturbation theory can be applied in this case.

The second issue complicating the problem is that the perturbative part of the
Hamiltonian generates series expansions whose radius of convergence does not
extend to the isotropic limit $\lambda=1$ but rather is limited to a circle
with a radius smaller than 1. Therefore, we analytically continue outside the
radius of convergence with the use of Pad\'e approximants. These are described
in detail in the literature \cite{Baker}. The algorithms used for their
calculation here are the determinant algorithm and the Viscovatov algorithm
\cite{Cuyt}. We now apply the method to two systems, $C_{12}$ and $C_{20}$.

\section{$C_{12}$}

A smaller system with a geometry similar to the ones considered above is the
truncated tetrahedron $C_{12}$ \cite{Trugman}, shown in figure 1. This is made
up of four triangles and three hexagons, and looks like a closed triangular
lattice. Every site is three fold coordinated, and bonds between sites on the
same triangle (tr.) are called $J_{1}$, while different triangle bonds (n.tr.)
are named $J_{2}$. The exact solution of the Heisenberg Hamiltonian has been
found for this system and is used to check the results from perturbation theory
here. At the classical level it was found to have a jump in the susceptibility
at a critical value of an external magnetic field and so has similar properties
to the systems discussed in reference \cite{Coffey}. As mentioned above the
classical solution determines local axes of quantization $z_{i}$ and determines
$H_{0}$. Spins belonging to the same triangle are coplanar at the classical
level and each local $x_{i}$ axis is defined in this plane. With application of
the right hand rule the corresponding $y_{i}$ axis is defined. With
$\alpha = \frac{J_{2}}{J_{1}}$ the Hamiltonian becomes:
\begin{equation}
 \frac{H}{J_{1}} = H_{0}+\lambda H_{1} \hspace{5pt},
\end{equation}
where
\begin{equation}
 H_{0} = \sum_{<i,j>}^{tr.} (-\frac{1}{2}) \ S_{i}^{z} S_{j}^{z} \ + \ \alpha
\sum_{<i,j>}^{n.tr.} (-1) \ S_{i}^{z} S_{j}^{z} \hspace{5pt},
\end{equation}
%\begin{equation}
% H_{1} = \sum_{<i,j>}^{tr.}\ (-\frac{3}{8}S_{i}^{+}S_{j}^{+} + \frac{1}{8}S_{i}^{+}S_{j}^{-}-\frac{\sqrt{3}}{4}S_{i}^{+}S_{j}^{z}+\frac{\sqrt{3}}{4}S_{j}^{+}S_{i}^{z}) + \alpha \sum_{<i,j>}^{n.tr.}\frac{1}{3}(\frac{1}{2}+\sqrt{2}i)S_{i}^{+}S_{j}^{+} + h.c. \hspace{5pt},
%\end{equation}
\begin{eqnarray}
\nonumber
H_{1}&=&\frac{1}{2} \ \sum_{<i,j>}^{tr.}\ ( \ \sin^{2}(\frac{\pi}{6}) \,
S_{i}^{+}S_{j}^{-} \, -\cos^{2}(\frac{\pi}{6}) \, S_{i}^{+}S_{j}^{+} \, - \,
\cos(\frac{\pi}{6}) \, S_{i}^{+}S_{j}^{z} \, + \, \cos(\frac{\pi}{6}) \,
S_{j}^{+}S_{i}^{z} \ ) \
\\ \cr
&+& \ \frac{\alpha}{2} \sum_{<i,j>}^{n.tr.} e^{i\phi} \,
S_{i}^{+}S_{j}^{+} \ + \ h.c.
\hspace{5pt},
\end{eqnarray}
where $\phi=\tan^{-1}(2\sqrt2$). The coefficients reflect the dependence of the
Hamiltonian on the classical spin directions. The Hamiltonian is complicated,
since it includes any possible combination of raising, lowering and $S_{i}^{z}$
operators, and has complex coefficients. There is no choice of the local
coordinate systems that would make all the coefficients real, once the local z
axes are fixed along the classical solution's directions. The local $x_{i}$ and
$y_{i}$ can also be defined without reference to the specific form of the
classical solution, as was stated in section II, but this results in a more
complicated expression for the Hamiltonian.

\subsection{Ground State Energy and Wavefunction}

The ground state energy and the wavefunction coefficients are functions of
$\lambda$ and $\alpha$. The $\alpha=0$ case corresponds to four isolated
triangles, while the $\alpha\to\infty$ case corresponds to spins forming
singlets (dimers) via the $J_{2}$ bond. There is no further frustration when
assembling the tetrahedron from the individual triangles, since this costs
nothing in energy for the classical spins. Consequently for this system the
classical ground state is independent of $\alpha$ and the quantum fluctuations
%introduce disorder at the $J_{2}=0$ and $J_{2}\to\infty$ limits \cite{Coffey}.
select a unique ground state when $\lambda \not= 0$.
This is reminiscent of the order-disorder transition induced by quantum
fluctuations for frustrated systems \cite{Millonas}. The effective Hamiltonian
for the two degenerate ground states has the following form:
%\begin{displaymath}
\begin{equation}
%\mathbf{H^{eff}} =
H^{eff} =
\left( \begin{array}{cc}
A_{N}(\lambda)&B_{N}(\lambda)\\
C_{N}(\lambda)&D_{N}(\lambda)
\end{array} \right) \hspace{5pt},
\end{equation}
%\end{displaymath}
where $A_{N}(\lambda)$, $B_{N}(\lambda)$, $C_{N}(\lambda)$ and $D_{N}(\lambda)$
are polynomials in $\lambda$ of Nth order, the order of perturbation. $\lambda$
can assume complex values, since we are also interested in the structure of the
functions in the complex plane. The coefficients of series $C_{N}(\lambda)$ and
$D_{N}(\lambda)$ are complex conjugates of the coefficients of $B_{N}(\lambda)$
and $A_{N}(\lambda)$ respectively. Therefore, in the case of physical interest
where $\lambda$ is real, $C_{N}(\lambda)$ and $D_{N}(\lambda)$ are the complex
conjugates of $B_{N}(\lambda)$ and $A_{N}(\lambda)$ respectively.
Diagonalization of the matrix in the latter case shows that its eigenvalues are
real, as expected. The two classical ground states, $|0\rangle$ and
$|\tilde{0}\rangle$, evolve in the following manner as functions of $\lambda$:
\begin{equation}
|\Psi_{0}\rangle_{N} = |0\rangle + \sum_{n=1}^{N} \lambda^{n} |\Psi_{n}\rangle
= |0\rangle + \sum_{i=1}^{d-L} F_{N}^{i}(\lambda)|i\rangle \hspace{5pt},
\end{equation}
and similarly for $|\Psi_{\tilde{0}}\rangle_{N}$, where $F_{N}^{i}(\lambda)$
are polynomials of Nth order in $\lambda$ with complex coefficients, $|i>$ is a
Hilbert space vector different from $|0\rangle$ and $|\tilde{0}\rangle$ and d
is the dimensionality of the Hilbert space. L is again the dimensionality of
the degenerate subspace (here $L=2$).

\subsubsection{Analytic Continuation}

The next step is to analytically continue the polynomials $A_{N}(\lambda)$,
$B_{N}(\lambda)$, $C_{N}(\lambda)$, $D_{N}(\lambda)$ and $F_{N}^{i}(\lambda)$.
Then the $2\times 2$ matrix is diagonalized and the energies as well as the
wavefunctions are known. This gives the ground and the first excited state at
each value of $\lambda$. For $C_{12}$ the dimensionality of the Hilbert space
is $d = 2^{12} = 4096$.

The ground state energy for different $\alpha$'s is given by the converged
value of $E_{N}^{g}(\lambda)$, which is the lowest eigenvalue for the $N$th
order approximant. The criterion for convergence is that at order N mean square
fluctuations should be $1\%$ of the mean value for the 7 approximants placed
around the Nth order (i.e.,$N-3,...,N,...,N+3$). This is easily satisfied for
small values of N when $\lambda<0.5$ for $\alpha=1$, but the mean square
fluctuations increase for larger values of $\alpha$. The dependence on $N$ for
different values of $\alpha$ can be understood by looking at the structure of
$A_{N}(\lambda)$ in the complex coupling constant plane (figures 2 and 3) which
is discussed below. The analytically continued form for the polynomials is
given as the ratio of two polynomials defined in the complex $\lambda$ plane so
that, for example:
\begin{equation}
A_{N}(\lambda) \to A_{N}^{a.c.}(\lambda) =
\frac{P_{N}(\lambda)}{Q_{N}(\lambda)} \hspace{5pt},
\end{equation}
where $a.c.$ stands for analytically continued. For $\lambda$ on the real axis
$D_{N}^{a.c.} = (A_{N}^{a.c.})^{*}$ and $C_{N}^{a.c.} = (B_{N}^{a.c.})^{*}$
guaranteeing real eigenvalues for real $\lambda$.
The structure is revealed by looking at zeros and poles of
$A_{N}^{a.c.}(\lambda)$ for complex $\lambda$. These are the roots of
$P_{N}(\lambda)$ and $Q_{N}(\lambda)$. Frustration leads to branch cuts in the
$\lambda$ plane which are given by lines of mixed zeros and poles of
$A_{N}^{a.c.}(\lambda)$. This is how the single valued $A_{N}^{a.c.}(\lambda)$
tries to reproduce a multi-valued function associated with a cut in the complex
plane. This structure is shown for $\alpha=1$ in figure 2 and for $\alpha=2$ in
figure 3, for $N=220$. The radius of convergence $\lambda_{c}$ (circles)
shrinks as $\alpha$ increases and for $\lambda>\lambda_{c}$ the perturbation
expansion is asymptotically convergent. This indicates that more orders are
needed for Pad\'{e} approximants to converge as the ratio
$\frac{\lambda}{\lambda_{c}}$ increases. One explanation for this is that, as
$\alpha$ increases, the classical ground state is not as good starting point as
for small $\lambda$, since the spins tend to form singlets via the $J_{2}$ bond
in the quantum limit. Thus it gets harder to reach the quantum state from the
ordered classical ground state, the former made up of independent singlets on
the non-triangle bonds at the limit $\frac{J_{2}}{J_{1}}\to\infty$.

% It is also
%pointed out that the triangles contain three spins, so the total spin can be as
%low as $\frac{1}{2}$, whereas in the $J_{2}$ bond case the spin is zero when
%singlets are formed. Therefore, the former case is closer to the classical
%state because the total spin is bigger, and when $J_{1}$ dominates $J_{2}$ we
%expect better convergence.

\subsubsection{Numerical Precision}

As we go to higher orders in perturbation theory the number of calculations
goes up, increasing the possibility for significant propagation of numerical
error \cite{Beach}. To extend the calculation to higher orders, the package
MPFUN was used \cite{MPFUN}, which allows arithmetic to very high precision,
limited only by machine specifications. The perturbation and analytic
continuation were done by using typically 94 digits precision in MPFUN, except
in some cases where 194 digits were used.

% The classical ground
%state for $C_{12}$ is analytically defined, but when it is not (as is the case
%in later sections), MPFUN accuracy was also used to calculate it.

To reduce
memory requirements and execution time, perturbation theory was first run in
double precision. After the states with equal coefficients or coefficients
differing only in sign (in real and/or imaginary part) due to symmetry were
identified, the program was run with MPFUN, taking advantage of these
symmetries. Thus the scale of the calculation was significantly reduced.
% Because of
%the complicated form of the Hamiltonian an analytic classification of the
%symmetries would be hard, therefore this more practical way was used instead.
The time required to get 250 orders in perturbation theory was approximately
thirty five minutes, when eight processors were used in parallel on a SGI
machine. These were IP27 processors with a frequency of 250 MHz. We typically
used 250 orders to get convergence for fields smaller than $J_{1}$.
%(even though this is probably not the optimal number of processors for this
%system).

A second criterion of convergence comes from the total spin component
$<S^{z}>$, which is a good quantum number for the Hamiltonian in the isotropic
case where $\lambda=1$. If the analytically continued wavefunction coefficients
converge, then the calculation of $<S^{z}>$ should yield an integer in an
applied field. Any deviation from an integer value indicates lack of
convergence, which could be due to insufficient orders of perturbation theory
used or propagation of numerical error, and indicates that higher numerical
accuracy is needed. This criterion is more stringent since now all the
coefficients $F_{N}^{i}(\lambda)$ have to converge but once convergence is
achieved any expectation value can be calculated, since the knowledge of the
wavefunction fully solves the problem. Since all the states in the Hilbert
space determined by the local z axes contribute to the ground and first excited
state coefficient functions, the calculated wavefunction is essentially exact.

\subsubsection{Degeneracy}

In the absence of an applied field states with different $<S^{z}>$ values are
degenerate and the excited state, a triplet in $C_{12}$, is a linear
combination of $S^{z}=0,\pm1$ states. However it is found that starting in the
$S^{z}=0$ sector as we do the perturbed state stays in this sector for all
$\lambda$ for both ground and excited state at h=0. Although this degeneracy is
lifted by the applied field, there is an additional degeneracy due to the
geometrical symmetries of $C_{12}$. The exact diagonalization shows that there
are three degenerate triplet states at the first excitation energy so that even
in an applied field the excited state with $S^{z}=1$ is triply degenerate.
Choosing a particular classical ground state to generate the starting states
$|0\rangle$ and $|\tilde{0}\rangle$ picks out a linear combination of these
degenerate states.

\subsubsection{Correlation Functions}

After the calculation of the ground state wavefunction, its correlation
functions can be directly evaluated. In $C_{12}$ there are five kinds of
qualitatively different correlation functions, two of which refer to nearest
neighbors. In figures 4 and 5 these are plotted as a function of the strength
of the quantum fluctuations $\lambda$ for the case $\alpha=1$ for the ground
and excited states respectively. The magnitude of the correlation function
$<\vec{S}_{1} \cdot \vec{S}_{9}>$ is smaller than $0.02$ and is not plotted.
For $\lambda=1$ the solution of the full isotropic case agrees with the one
found from exact diagonalization \cite{Trugman}. The nearest neighbor
correlation functions are $<\vec{S}_{1} \cdot \vec{S}_{2}> = -0.125$ and
$<\vec{S}_{1} \cdot \vec{S}_{4}> = -0.250$ at the classical level where
$\lambda=0$. Spins 1 and 4 are antiparallel, while
$<\vec{S}_{1} \cdot \vec{S}_{2}>$ is one half of
$<\vec{S}_{1} \cdot \vec{S}_{4}>$.

In the ground state (figure 4) $<\vec{S}_{1} \cdot \vec{S}_{2}>$ increases in
magnitude with $\lambda$ and at $\lambda=1$ equals $-0.183$, being roughly
$50\%$ bigger than its classical value. $<\vec{S}_{1} \cdot \vec{S}_{4}>$
increases in magnitude as the quantum fluctuations become stronger and reaches
its maximum just before $\lambda=0.8$. Then it doesn't change significantly and
its magnitude at $\lambda=1$ is more than two times larger than its classical
one, being equal to $-0.586$. This is because as the perturbation is turned on
adjacent spins belonging to different triangles want to create singlet bonds
due to their $J_{2}$ interaction. The value of
$<\vec{S}_{1} \cdot \vec{S}_{4}>$ at $\lambda=1$ is close to the value for a
singlet state between two spins, which is $-0.750$. The other two correlation
functions have a non-trivial dependence on $\lambda$, but their values in the
isotropic case do not significantly differ from their classical ones.

The excited state correlation functions of figure 5 do not differ significantly
from the ones of figure 4.
%This is due to the fact that the excited state
%belongs to the triplet for which $S=1$ but its $S^{z}=0$, due to the time
%reversal symmetry of the Hamiltonian. Consequently, $S^{z}=0$ for any value of
%$\lambda$, as is also the case for the singlet ground state.
However, we observe a smaller value for $<\vec{S}_{1} \cdot \vec{S}_{5}>$
compared with the ground state value, while $<\vec{S}_{1} \cdot \vec{S}_{4}>$
has a smaller magnitude at $\lambda=1$ again compared to the ground state value
indicating a more triplet character for this bond.

In figures 6 and 7 we plot the correlation functions for $\alpha=2$. We observe
that for both the ground and the excited state all the correlations decrease in
magnitude, except the ones between neighboring spins connected via the $J_{2}$
bond. The value of $<\vec{S}_{1} \cdot \vec{S}_{4}>$ in the ground state is now
$-0.697$, approaching the singlet value $-0.750$ even closer.

%\subsection{Excited States}
%The states with one spin flipped at the classical level are the elementary
%excitations of the classical Hamiltonian. We can do perturbation theory on this
%degenerate subspace and recover the excitation spectrum in the quantum case.
%Now the dimensionality L of the degenerate subspace, in the case where
%$\alpha=1$, is equal to 36. The amount of computer memory required for a
%numerical approach is now bigger.

\subsection{Magnetic Field}

We next introduce a magnetic field in the problem and study the ground and
excited states as a function of $\lambda$. There are two possible approaches to
this calculation. The first is to perturb around the zero magnetic field
classical ground state. The second is to calculate the classical ground state
in the presence of a magnetic field, and then apply perturbation theory.
Because the field breaks the time reversal symmetry, the classical ground state
is non-degenerate, and in the second case non-degenerate perturbation theory is
used. The results of the two methods should agree at the isotropic limit,
$\lambda=1$. However, their dependence on $\lambda$ is different.

\subsubsection{Degenerate case. Field independent classical ground state.}

The Hamiltonian for the first method has the form:
\begin{equation}
 \frac{H}{J_{1}} = H_{0}+\lambda ( H_{1} + H_{2} ) \hspace{5pt},
\end{equation}
where $H_{0}$ and $H_{1}$ were defined before and $H_{2}$ is the part that
relates to the magnetic field. Specifically:
\begin{equation}
 H_{2} = - h \sum_{i=1}^{N} cos\theta_{i} S_{i}^{z} + \frac{h}{2}
         \sum_{i=1}^{N} sin\theta_{i} (S_{i}^{+}+ S_{i}^{-}) \hspace{5pt}.
\end{equation}
h is the strength of the magnetic field in units of $J_{1}$ and $\theta_{i}$
the classical solution's angles with the global z axis. The direction of the
magnetic field is taken along the global z axis, so it is perpendicular to the
plane of one of the four triangles of the system. The spins belonging to any
triangle are coplanar in the classical case (in spin space), and they can be
chosen to lie in the physical plane of the triangle.

\subsubsection{$<S^{z}>$}

For $\lambda=1$, the ground state lies in the $S^{z}=0$ sector when there is no
magnetic field. $<S^{z}>$ is defined along the global z axis, and it commutes
with the Hamiltonian (13) at $\lambda=1$. As the field is turned on the energy
of the ground state, which is a singlet, won't change at the isotropic limit.
However, the triplet excited state has $S^{z}=1$ there, and its energy
decreases linearly with the magnetic field due to the Zeeman term. At a
critical value of the field the triplet state energy becomes equal to the one
of the singlet state, so the triplet state becomes the ground state. As the
magnetic field increases further the ground state moves towards spin sectors
with higher value of $S^{z}$, until the magnetization saturates. The results
for the energies found from perturbation theory and analytic continuation are
found in figure 8, and they reproduce the exact values ($\lambda=1$). By
extrapolation of the straight lines which give the ground and the first excited
energy back to zero field, we recover the energies of excited states in the
zero field case. The diagonalization of the effective Hamiltonian matrix gives
the following eigenenergies:
\begin{equation}
E_{1,2} = \frac{A_{N}^{a.c.}(\lambda)+D_{N}^{a.c.}(\lambda)}{2} \mp \Bigg \arrowvert
\sqrt{B_{N}^{a.c.}(\lambda)C_{N}^{a.c.}(\lambda) +
(\frac{A_{N}^{a.c.}(\lambda)-D_{N}^{a.c.}(\lambda)}{2})^2} \Bigg \arrowvert
\hspace{5pt},
\end{equation}
where $A_{N}^{a.c.}(\lambda)$, $B_{N}^{a.c.}(\lambda)$, $C_{N}^{a.c.}(\lambda)$
and $D_{N}^{a.c.}(\lambda)$ are the analytically continued values of the
corresponding polynomials of $Nth$ order. At the critical fields at which the
two states exchange 'roles' as the ground and the first excited state, the
square root goes to zero at $\lambda=1$. This occurs when $B_{N}^{a.c.}$ and
the imaginary part of $A_{N}^{a.c.}$ go to zero. In this way, we start from two
degenerate ground states and we end up with two degenerate states.

The strict criterion for successful convergence of the wavefunctions is the
calculation of the $<S^{z}>$ number. This should be an integer at $\lambda=1$.
We calculate its evolution with $\lambda$ in the global spin basis, where the
quantization axis is the same for all spins. The expression for its expectation
value $<S^{z}>$ is:
\begin{equation}
 <S^{z}> = \frac{<\Psi_{0}|S^{z}|\Psi_{0}> + |G|^{2}
           <\Psi_{\tilde{0}}|S^{z}|\Psi_{\tilde{0}}> +
           2Re(G<\Psi_{0}|S^{z}|\Psi_{\tilde{0}}>)}
           {<\Psi_{0}|\Psi_{0}> + |G|^{2} <\Psi_{\tilde{0}}|\Psi_{\tilde{0}}> +
           2Re(G<\Psi_{0}|\Psi_{\tilde{0}}>)} \hspace{5pt},
\end{equation}
where G is given by
\begin{equation}
G = \frac{\frac{A_{N}^{a.c.}(\lambda)-D_{N}^{a.c.}(\lambda)}{2} \pm
    \sqrt{B_{N}^{a.c.}(\lambda)C_{N}^{a.c.}(\lambda) +
    (\frac{A_{N}^{a.c.}(\lambda)-D_{N}^{a.c.}(\lambda)}{2})^2}}
    {B_{N}^{a.c.}(\lambda)} \hspace{5pt}.
\end{equation}
The plus sign corresponds to the ground state and the minus sign to the excited
one. The magnitude of this coefficient is $1$ for any real $\lambda$, since in
this case $D_{N}^{a.c.} = (A_{N}^{a.c.})^{*}$ and
$C_{N}^{a.c.} = (B_{N}^{a.c.})^{*}$. The accurate calculation of the
wavefunctions at the critical fields $h_{c}$ involves taking the limit
$\lim_{\lambda\to1}G(\lambda)$. In this case the magnitude of
$B_{N}^{a.c.}(\lambda)$ as well as the imaginary part of
$A_{N}^{a.c.}(\lambda)$ go to zero, thus both the numerator and denominator
vanish. The success of the analytic continuation is such that the calculation
of $A_{N}^{a.c.}(\lambda)$ and $B_{N}^{a.c.}(\lambda)$ is so accurate after
analytic continuation that the result $\lim_{\lambda\to1}G(\lambda)=\pm1$ is
recovered and this is reflected in the calculated $<S^{z}>$. The calculation
was done for various fields for $\alpha=1$, and the results are shown in
figures 9 and 10. There it is shown that indeed $<S^{z}>$ assumes integer
values when $\lambda=1$, confirming the success of analytic continuation. For
example, when $h=0.7$, at $\lambda=1$ $<S^{z}>=0.99999999998$ for the ground
state and $<S^{z}>=1.0 \times 10^{-11}$ for the excited state.

As seen in figure 9, the quantum fluctuations raise the value of $<S^{z}>$ for
small values of the magnetic field, but eventually it goes to zero at the
isotropic limit. However, just above the transition to the $S^{z}=1$ sector,
which occurs at $h_{c1}=0.6878$ in agreement with the exact answer, $<S^{z}>$
has a non-trivial behavior for intermediate $\lambda$'s before assuming the
value equal to $1$. Thus for $h=0.688$ there is a rapid change in the
magnetization at a value of $\lambda$ close to $1$. There is competition
between the $0$ and $1$ spin sector for the ground state, and finally the
quantum fluctuations lead to the latter. The magnetic field terms of the
perturbing Hamiltonian favor a magnetized ground state, while the rest favor
zero spin. The sudden change in $<S^{z}>$ as a function of $\lambda$ moves
closer to $\lambda=1$ as $h \to h_{c}$ from below. If we go further away from
this critical field, the jump is pushed towards smaller values of $\lambda$,
and eventually it vanishes. Similar effects are observed in figure 10, which
includes the corresponding graphs for the excited state. In this case, the
excited state has $S^{z}=1$ below and $S^{z}=0$ above the transition. The
conclusion is that there is a 'window' around the critical field where the
$\lambda$ dependence of $<S^{z}>$ is very strong.

As the magnetic field gets bigger, more terms are needed to analytically
continue the wavefunction so that the spin number assumes the proper integer
values. There is also a need for greater numerical precision, due to the
increased number of calculations which tend to propagate the numerical errors
\cite{Beach}, and so 194 digits of MPFUN accuracy were used. The transition
between the $S^{z}=1$ and $S^{z}=2$ spin sectors for the ground state takes
place at $h_{c2}=0.9869$. For a magnetic field equal to 1.01, 501 orders were
generated to get convergence. This requirement for more orders in perturbation
theory and more accuracy makes the calculation of the factor $G$ of equation
(16) harder as the second critical field is approached. The behavior of
$<S^{z}>$ as a function of $\lambda$ is shown in figure 11, where the $<S^{z}>$
value of the ground state remains constant over a range of $\lambda$, and
eventually ``jumps'' to the final value $<S^{z}>=2$. On the other hand, the
excited state $<S^{z}>$ approaches 2 close to $\lambda \approx 1$ only to
settle at 1 when $\lambda=1$.

%The analytic continuation of the wavefunction didn't
%work well close to $\lambda=1$, even though it worked fine just below and just
%above this value. This is again indicative of the problem of the division of
%very small numbers to find $G$, which was mentioned before and becomes more
%severe very close to the critical fields. However, the calculation of the
%energies is not affected at all since there is no division that has to be done
%to calculate them as seen in equation (14).

Non-monotonic behavior of the quantum number $S^{z}$ near a transition is
observed for other values of $J_{2}$ as well. These are shown in figures 12 and
13, for $J_{2}$ values equal to $0.8$ and $1.8$. For the second case the higher
value of $J_{2}$ stabilizes the spin value for $\lambda's$ close to $1$. It was
observed that convergence was harder to get as $J_{2}$ and/or $h$ were
increased, for the reasons already mentioned.

\subsubsection{Correlation Functions}
%The two point correlation functions were also calculated for a magnetic field
%$h=0.7$. They are plotted in figures 13 and 14 for the ground state, a triplet,
%and the excited state, a singlet, respectively. Correlation function
%$<\vec{S}_{1} \cdot \vec{S}_{4}>$ has a non-trivial dependence in $\lambda$ for
%both of them. The minimum for the ground state and the saddle point for the
%excited state occur at $\lambda \approx 0.5$ where the $<S^{z}>$ (plotted in
%figures 8 and 9) also rapidly changes value towards its $\lambda=1$ value.

%In
%the ground state (figure 13), which is the excited state at $h=0$ (figure 4),
%the magnetic field doesn't favor the $J_{2}$ bond nearest neighbor correlation,
%and this has decreased. On the other hand, in the singlet excited state (figure
%14), which was the ground state at $h=0$ (figure 3), everything is the same at
%$\lambda=1$. The quantum fluctuations are favoring a field-ordered state as the
%magnetic field inreases. This can be seen by comparing the values at
%$\lambda<1$, and they are smaller in the magnetic field.

%As a further check on the accuracy of the wavefunctions we calculate
%$<\vec{S}^{2}>$ in the ground and excited states. These are $-0.0028361$ for
%the $S=0$ state and $2.00188$ for $S=1$, consistent with the value $S(S+1)$.

Although the calculated eigenstates at $\lambda=1$ are independent of magnetic
field they do depend on the field for intermediate values of $\lambda$. This
can be seen clearly in $<S^{z}>$ as a function of $\lambda$ in figures 9 and
10. As pointed out previously the choice of the starting classical ground state
picks out a particular linear combination of the three degenerate ($S=1$,
$S^{z}=1$) states. Different linear combinations lead to different values of
the $<\vec{S}_{i} \cdot \vec{S}_{j}>$. However when adding the values of
nearest neighbor $<\vec{S}_{i} \cdot \vec{S}_{j}>$ for different choices of
starting classical ground states the values of the energies, calculated
directly in equation (15), are recovered.

A further check on the two point correlation functions calculated for the
wavefunctions is $<\vec{S}^{2}> = \sum_{i,j} <\vec{S}_{i} \cdot \vec{S}_{j}>$.
Calculating $<\vec{S}^{2}>$ in the ground and excited state at $h=0.7$ we find
$2.00188$ and $-0.0028361$ consistent with $<{S}^{2}>=S(S+1)$ for $S=1$ and
$S=0$. This measurement tests the accuracy of all two point correlation
functions. The ones for nearest neighbors for $h=0.7$ are plotted in figures 14
and 15 for the ground and excited states. Comparing these with the $\lambda$
dependence of the states in figures 6 and 7, the only difference is a
non-monotonic dependence at $\lambda \simeq 0.5$ especially for the nearest
neighbor correlation functions $<\vec{S}_{1} \cdot \vec{S}_{2}>$ and
$<\vec{S}_{1} \cdot \vec{S}_{4}>$.

\subsubsection{Analytic Structure}

The motion of the zero of the square root appearing in equations (15) and (17)
in the complex $\lambda$ plane can be traced out with the help of the Pad\'e
approximants, which are ratios of two polynomials. Since the analytic
continuation was quite successful in the vicinity of $\lambda=1$, we can look
for the structure of approximants in this area. To this end, we consider the
function
\begin{equation}
B_{N}^{a.c.}(\lambda)C_{N}^{a.c.}(\lambda) +
\Bigg(\frac{A_{N}^{a.c.}(\lambda)-D_{N}^{a.c.}(\lambda)}{2}\Bigg)^2
\hspace{5pt},
\end{equation}
which is inside the square root in equations (15) and (17), and we plot the
zeros of the numerator and the denominator of the approximants as a function of
the magnetic field. The approximants were generated
%used are the ones generated with the Viscovatov algorithm
%('staircase' approximants) after subtraction of the zero terms of the leading
%orders \cite{Cuyt}, and
with 220 orders of perturbation theory (the order
of the numerator and the denominator of the Pad\'e approximant is one half of
this number). In figure 16 the complex plane structure is plotted for a
magnetic field h=0.688. The Pad\'e approximant is seen to reproduce the branch
cuts of the function in the complex plane. There are also zeros and poles in
the complex plane not falling on any of the branch cuts. These are present only
in approximants of specific order and they are artifacts of the analytic
continuation. It can be seen how the zeros of equation (18) appear in the
complex $\lambda$ plane close to $\lambda=1$. The coefficients of the expansion
of (18) are real, so the zeros and the poles appear in complex conjugate pairs.

Another feature of the picture is that it is asymmetric with respect to the
imaginary $\lambda$ axis. This asymmetry gets more and more pronounced with
increasing field. This is because negative $\lambda's$ correspond to different
ferromagnetic couplings between neighboring sites, and these interactions favor
aligned spins, a state similar to the one favored by the magnetic field. In
this case it is more difficult to get convergence because of the competition of
the classical antiferromagnetic interactions with the quantum ferromagnetic
ones and the magnetic field. Consequently the radius of convergence is smaller.

%and such a state would be very hard to reproduce here where the starting point
%is a state antiferromagnetically ordered along the spin local axes. In other
%words, when $\lambda$ is substituted by $-\lambda$ it is as if we change the
%sign of the interaction along the local z axes. Consequently, the classical
%ground state becomes the ferromagnetic one, and due to this some spin axes
%would have to be flipped. If not, as is the case here, the commutation
%relations between spin components where the spin z component is involved will
%change sign.

In figure 17 the roots of the approximants are plotted for various magnetic
fields close to $h_{c1}=0.6878$ in the vicinity of $\lambda=1$. The zeros of
the square root approach $\lambda=1$ in conjugate pairs and they finally hit
the axis at the transition. As the strength of the field is further increased,
the roots move away from $\lambda=1$ in the opposite direction. We see this
behavior in a small range of the fields around the transition with roots around
$\lambda=1$, which is reflected in the rapid non-monotonic variation of
$<S^{z}>$ around the critical field. Since the square root of function (18) has
to eventually be taken, its zeros will turn into branch points in the complex
$\lambda$ plane.

%As the field is increased, more and more terms are required for convergence,
%since the radius of convergence of the series expansions shrinks because the
%perturbation is getting stronger. This can be seen by plotting the zeros of the
%function $A_{N}$ mentioned in the previous section (figure 15). This is done
%for $h=0$ and $h=0.8$, which correspond to a ground state in the $S^{z}=0$ and
%$S^{z}=1$ sectors respectively. The function is seen to have a bigger radius of
%convergence when $h=0$.

%The transition from the $S^{z}=1$ to the $S^{z}=2$ sector takes place at
%$h=0.9869$. We examine the $\lambda$ dependence of $S^{z}$ in this area. This
%is shown in figure 15. For fields above and below the transition, there is a
%plateau and then the spin settles to its final value. The two spin sectors are
%competing for the ground state, and only when $\lambda$ gets close to 1 of the
%two sectors is picked up. This is similar to what happens in the 0 to 1
%transition.

\subsubsection{Non-degenerate case. Field dependent classical ground state.}

 The alternative approach to the problem is to calculate the classical ground
state in the presence of the magnetic field, and then introduce the remaining
terms as perturbation. In this case the Hamiltonian is:
\begin{equation}
 \frac{H}{J_{1}} = H_{0}+H_{0}^{'}+\lambda ( H_{1} + H_{2}^{'} ) \hspace{5pt},
\end{equation}
where:
\begin{equation}
 H_{0}^{'} = - h \sum_{i=1}^{N} cos\theta_{i} S_{i}^{z} \hspace{5pt},
\end{equation}
\begin{equation}
 H_{2}^{'} = \frac{h}{2} \sum_{i=1}^{N} sin\theta_{i} (S_{i}^{+}+ S_{i}^{-})
\hspace{5pt}.
\end{equation}
and $H_{0}$ and $H_{1}$ have already been defined in equations (8) and (9).
Since the magnetic field breaks the degeneracy of the classical ground state
Rayleigh-Schr\"odinger perturbation theory can be applied directly for the
wavefunctions. The results after the analytic continuation are shown in figure
18. This method can be compared with the one of the previous section only at
$\lambda=1$, where in both cases the isotropic Heisenberg Hamiltonian is
recovered. Starting from the field dependent classical ground state is not any
better as far as computer memory and execution time is concerned, since
symmetries reduce memory requirements at the same level in both cases. However,
the first method also gives the ground and first excited state very accurately,
consequently the approach of the transitions between $S^{z}$ sectors as a
function of the applied magnetic field is easily seen. On the other hand, the
second method works better for higher fields, since now the classical ground
state is related to the strength of the magnetic field, thus it has a non-zero
value for $<S^{z}>$. Various fields up to $h=1.3$ where employed in the
calculation. The results for the energy agree with the exact diagonalization
answer \cite{Trugman}.
%For the spin number there was not convergence at $\lambda=1$ at the expected
%integer values for some fields, indicating problems with the analytic
%continuation of the wavefunction coefficients.

Looking at figure 18, we see the dependence of $<S^{z}>$ on $\lambda$ is not
monotonic for the various magnetic fields. This can be attributed to the fact
that the starting point is now the magnetic field dependent classical ground
state. The terms in the Hamiltonian related to the magnetic field tend to
increase the spin, while the rest favor a zero spin quantum state, thus there
is competition between the two at the classical and quantum level. In the
degenerate case the field dependent terms entered only in the perturbing part
of the Hamiltonian, making the perturbation stronger and dominant in the
determination of $S^{z}$ for higher $\lambda$'s.

The structure of the analytic continuation of the energy was also studied in
the non-degenerate case, and it was consistent with the structure found from
the degenerate perturbation. For a magnetic field $h=0.8$ the expression in
equation (17) has a zero at $\lambda_{1} \approx 0.788+0.337i$, while the
analytically continued function for the energy in the non-degenerate case has a
branch cut which starts at a value of $\lambda$ equal to
$\lambda_{2} \approx 0.898+0.217i$ and extends almost parallel to the x axis.
Since $\sqrt{\mid\lambda_{1}\mid}\simeq{\mid\lambda_{2}\mid}$, the branch cut
of the square root in the degenerate case corresponds to the one of the
non-degenerate case demonstrating the consistency of the two approaches
starting from the classical ground state in the absence of a magnetic field and
the magnetic field dependent classical ground state.

\section{$C_{20}$}

Next we consider the 20 site system on the vertices of a dodecahedron, shown in
figure 19. In $C_{20}$ there is only one kind of bond, which we call J, and
each atom has three nearest neighbors as in the $C_{12}$ case. In the classical
solution spins belonging to the same pentagon are not coplanar \cite{Coffey},
and the energy per bond is $-\frac{\sqrt{5}}{3}=-0.7454$. This is bigger than
the energy per bond for the coplanar spins on an isolated pentagon, which is
%$cos(\frac{4\pi}{5})=-\frac{\sqrt{5}+1}{4}=-0.8090$.
$cos(\frac{4\pi}{5})=-0.8090$. Thus when the dodecahedron is assembled from the
individual pentagons there is a cost in energy, in contrast to the twelve-site
system case. This system has a discontinuity in $<S^{z}>$ as a function of the
magnetic field at the classical level unlike the twelve-site system. The
discontinuity has a magnitude of $0.64$ and occurs at a magnetic field
$h=1.432J$. This is also the case for $C_{60}$ whereas in $C_{12}$, $C_{70}$
and $C_{84}$ there is a discontinuity in the slope of the magnetization with
applied field \cite{Coffey}. We start again from the classical ground states
and perturb them with the quantum fluctuations. The difference now is that the
exact solution of the problem is not known and the Hilbert space is much larger
with $2^{20}=1,048,576$ states.

The question we wish to address is whether the differences between $C_{12}$ and
$C_{20}$ seen at the classical level in magnetic properties survive for the
$S=\frac{1}{2}$ case. As we will show below these differences do survive and
are seen in the analytic structure of the perturbation theory for the two
systems. Since spins are not coplanar at the classical level, the Hamiltonian,
which is defined as described in section II, assumes the following complicated
form:

\begin{equation}
     \frac{H}{J} = H_{0}+\lambda H_{1} \hspace{5pt},
\end{equation}

\begin{equation}
 H_{0} = -\frac{\sqrt{5}}{3} \sum_{<i,j>} \ S_{i}^{z} S_{j}^{z} \hspace{5pt},
\end{equation}
\begin{equation}
 H_{1} = \sum_{<i,j>}\ (\alpha_{ij}S_{i}^{+}S_{j}^{+} + \beta_{ij}S_{i}^{+}S_{j}^{-}+\gamma_{ij}S_{i}^{+}S_{j}^{z}+\delta_{ij}S_{j}^{+}S_{i}^{z}) + h.c. \hspace{5pt},
\end{equation}

where $\alpha_{ij}$, $\beta_{ij}$, $\gamma_{ij}$ and $\delta_{ij}$ are complex
coefficients defined analytically. For example, for the
$\vec{S}_{1} \cdot \vec{S}_{2}$ term
$\alpha_{12}=\frac{1}{4}(1+\frac{\sqrt5}{3})$,
$\beta_{12}=\frac{1}{8}(1-\frac{\sqrt5}{3}) (1-\sqrt3i)$,
$\gamma_{12}=-\frac{1}{6}(1+\sqrt3i)$ and $\delta_{12}=\gamma_{12}^{*}$. Since
here the starting point is the classical ground state, defining a local z axis
at each site, the coefficients of $H_{1}$ are in general different for
different bonds. Furthermore they are complex because of the non-planar
character of the spins in the classical ground state. We apply the same methods
towards the solution as in the twelve-site system case. The elements of the
$2\times2$ effective Hamiltonian matrix $H^{eff}$ in equation (10) are now real
for all applied magnetic fields.

\subsection{Ground and Excited States}

The energies and wavefunctions of the two lowest lying states were calculated
up to $h=0.72J$. The ground state ($\circ$) and the excited state ($\diamond$)
are shown in figure 20. By drawing straight lines through the energies and
extrapolating to zero field the lowest energy state in each spin sector is
recovered, as well as the second lowest energy in the $S=0$ sector. We find
these energies to be $E_{1}(S=0)=-9.722J$, $E_{2}(S=0)=-9.345J$,
$E_{3}(S=1)=-9.208J$ and $E_{4}(S=2)=-8.523J$.
%Their magnetic field dependence
%is given by $E_{i}=E_{i}(h=0) - S^{z}h$ yielding transitions at $h=0.147$
%between the second lowest singlet state ($S=0$) and the triplet state ($S=1$),
%$h=0.514$ between the lowest singlet state ($S=0$) and the lowest triplet state
%($S=1$), $h=0.600$ between the lowest singlet state ($S=0$) and the $S=2$
%state, and $h=0.683$ between the lowest states with $S=1$ and $S=2$.
The magnetic field dependence is given by a Zeeman term
$-h\sum_{i=1}^{20}S^{z}_{i}$. Unlike $C_{12}$ the first excited state at
$h\le0.137$ is a singlet rather than a triplet \cite{Mambrini}.

%The difference in the behavior of the two systems is attributed to their
%difference in the connectivity. Each site is three-fold coordinated in both
%systems and the average energies per nearest neighbor bond for $S=\frac{1}{2}$
%spins, $-0.325J$ (in $C_{20}$) and $-0.3105J$ (in $C_{12}$), taking
%$J_{1}=J_{2}=J$, are comparable. The difference in the character of the energy
%spectra of the two systems arises from the presence of an almost singlet
%non-triangle bond in the ground state of $C_{12}$. The first excited arises by
%taking a linear combination of states each associated with a triplet excitation
%of a single non-triangle bond. By comparing $<\vec{S}_{1} \cdot \vec{S}_{4}>$
%in figures 13 and 14 the non-triangle bond is seen to take on a more strongly
%triplet character in the excited state. On the other hand in $C_{20}$ the
%difference in $<\vec{S}_{1} \cdot \vec{S}_{2}>$ changes very little, $-0.32$ to
%$-0.30$ between ground and excited state.

%It is also observed that when the field has zero or a small value the diagonal
%and the off-diagonal terms are asymptotically equal in magnitude and opposite
%in sign.

%In figure 19 we plot the lambda dependence of the expectation value $<S^{z}>$
%for $h=0.4$. We see a behavior similar to the one of $C_{12}$, as the ground
%state assumes the value $0$ and the excited state the value $1$ at $\lambda=1$.
%Comparing with figure 21, we see how the correlations of the spins with the
%magnetic field leave their signature in the evolution of $<S^{z}>$ with
%$\lambda$.

\subsection{Correlation Functions}

%The ground and first excited state energy and wavefunction were calculated for
%a magnetic field $h=0.4$. These are the lowest lying singlet state and the
%$S^{z}=1$ state of the lowest lying triplet at $h=0$. There are 3 different
%nearest neighbor correlation functions because the local spin basis is used so
%that the geometrical symmetry of the system is broken, and in this basis the
%terms of the Hamiltonian are not the same for every pair of spins except at
%$\lambda=0$ and at $\lambda=1$ where the rotational symmetry of the Hamiltonian
%is recovered. However, all three of them are qualitatively similar.
%$<\vec{S_{1}} \cdot \vec{S_{2}}>$ and $<\vec{S_{1}} \cdot \vec{S_{6}}>$ are
%plotted in figure 20. The third nearest neighbor correlation function,
%$<\vec{S_{6}} \cdot \vec{S_{7}}>$, can not be distinguished in the scale used
%from $<\vec{S_{1}} \cdot \vec{S_{2}}>$. The same two point correlation
%functions are plotted in figure 21 for the excited state. The rest of the
%correlation functions involving $S_{1}$ are also plotted in figures 20 and 21
%for the ground and excited state respectively. Again, only the ones
%qualitatively different from each other are plotted.

In figure 21 we plot the lambda dependence of the expectation value $<S^{z}>$
for $h=0.4$. We see a behavior similar to the one of $C_{12}$, as the ground
state assumes the value $-2.7 \times 10^{-8}$ and the excited state the value
$0.9999997$ at $\lambda=1$. Again this points to the accuracy of the calculated
wavefunctions. The ferromagnetic or antiferromagnetic character of the
correlation between any two sites remains the same in the ground state and the
two lowest lying excited states at $h=0$. In each of these states the nearest
neighbor correlation $\sim -0.3$, while all other
$<\vec{S_{i}} \cdot \vec{S_{j}}>$ have magnitudes $\sim 0.03 \to 0.08$ at
$\lambda=1$. Correlation functions are plotted for the lowest lying singlet and
triplet states at $h=0.4$, where they are the ground and first excited state in
figures 22 and 23. We first consider the nearest neighbor correlations. For the
singlet state, figure 22, the $<\vec{S_{1}} \cdot \vec{S_{2}}>$ and
$<\vec{S_{1}} \cdot \vec{S_{6}}>$ are equal only at $\lambda=0$ and
$\lambda=1$. The $\lambda$ dependence for $0<\lambda<1$ depends on both the
starting classical ground state and the value of $h$. For the lowest energy
singlet state, $<\vec{S_{1}} \cdot \vec{S_{2}}>$ and
$<\vec{S_{1}} \cdot \vec{S_{6}}>$ grow in magnitude from their classical value,
$-0.186$ at $\lambda=0$, to $-0.324$ at $\lambda=1$ for all nearest neighbor
correlations. Thus they are comparable to the value of the nearest neighbor
correlation in the ground state of an isolated pentagon, $-0.375$.

In the lowest lying $S=1$ $S^{z}=1$ state, figure 23, two values of the nearest
neighbor correlation functions are present, $-0.297$ and $-0.327$ at
$\lambda=1$. This state can be described by considering the bonds on a
pentagon, on its diametrically opposite mirror and on a chain formed by the 10
remaining sites. On these bonds $<\vec{S_{i}} \cdot \vec{S_{j}}> = -0.297$
while on the bonds connecting the chain to the two pentagons
$<\vec{S_{i}} \cdot \vec{S_{j}}> = -0.327$. Since this configuration can be
chosen in six different ways the triplet $S=1$ $S^{z}=1$ state is six fold
degenerate. There is little difference between the next nearest neighbors and
more distant correlations in the singlet and triplet states. The quantum
fluctuations significantly reduce the magnitude of these correlations compared
with their classical values at $\lambda=0$.

As a further check of the accuracy of the wavefunction the calculated values of
these correlations give for
$<\vec{S}^{2}> = \sum_{i,j} <\vec{S}_{i} \cdot \vec{S}_{j}>$ the value
$-0.00089$ for the ground state and the value $1.99154$ for the excited state
at $\lambda=1$. These are consistent with the calculated $<S^{z}>$ values in
figure 21. The change in the value of $<\vec{S}^{2}>$ between the singlet and
triplet states comes from changes in all the $<\vec{S_{i}} \cdot \vec{S_{j}}>$.
The change due to the nearest neighbor $<\vec{S_{i}} \cdot \vec{S_{j}}>$'s is
$\sim 1$ and the rest comes from small changes, $\sim 0.005$, from the other
bonds. This suggests that there is no simple characterization of the excited
states.

%Examining the $<\vec{S}_{i} \cdot \vec{S}_{j}>$ at $\lambda=0$ and
%$\lambda=1$ one sees that the quantum fluctuations do not significantly reduce
%the effect of frustration on the nearest-neighbor bonds. The magnitude of
%$<\vec{S}_{1} \cdot \vec{S}_{2}>$ remains considerably smaller than its value
%on an isolated pentagon. The next nearest correlations
%$<\vec{S}_{1} \cdot \vec{S}_{3}>$ and $<\vec{S}_{1} \cdot \vec{S}_{9}>$ are
%reduced from their small classical values. Longer range correlations are
%greatly reduced by fluctuations.

%The most significant difference between ground and excited state wavefunctions
%is in the nearest-neighbor correlation functions where
%$<\vec{S}_{1} \cdot \vec{S}_{6}>$ is the same in both states but
%$<\vec{S}_{1} \cdot \vec{S}_{2}> \approx -0.3$ in the excited state and
%$\approx -0.35$ in the ground state.

In the classical ground state all the spins in the top and bottom pentagons
have the same azimuthal angle $\alpha$ with the z axis, which is later taken to
be the direction of the magnetic field. The rest of the spins form a different
angle $\beta$ with the z axis, so there are two different angles in the
classical solution. There are only two distinct functions
$<\vec{S}_{i} \cdot \vec{h}>$ of the spins for the calculated states in the
magnetic field, and they represent these two kinds of sites in the classical
ground state. They were calculated for $i=1$ and $6$ and they are plotted in
figure 24 for the ground and the excited state. They are consistent with
$<\sum_{i}\vec{S}_{i} \cdot \vec{h}> = <S^{z}>h$ calculated in the ground and
excited state in figure 21.

%Since the ground state is a singlet, the
%correlation functions with the field vanish at $\lambda=1$. Figure 21 is in
%agreement with figure 18, since the correlations of the spins with the field
%when added together give the value of $<S^{z}>$ scaled with the magnetic field.

For $h\le0.137$ the excited state is a singlet. This state has the same
symmetry as in the triplet state with two types of nearest neighbor bonds equal
to $-0.304$ and $-0.324$. As in the triplet case the longer range correlations,
both ferromagnetic and antiferromagnetic are still comparable in magnitude and
$\sim 0.05$.

\subsection{Analytic Structure}

Next we study the structure of the analytic continuation of the off-diagonal
element $B_{N}^{a.c.}(\lambda)$. In figure 25 we plot the structure at a
magnetic field equal to 0.3. Again we observe the asymmetry with respect to the
y axis.
%Then we look
%for the signature of the transition taking place at h=0.514 in the complex
%$\lambda$ plane. In figure 24 we plot the structure around $\lambda=1$ for a
%field just below and one just above the transition. We see how the zero of the
%off-diagonal element moves towards $\lambda=1$ as the critical field is
%approached and then the magnetic field increases above its value. The zeros are
%now along the real axis, in contrast with the $C_{12}$ case. This is due to the
%connectivity of the two systems, since now in $C_{20}$ there is only one kind
%of bond and the effective Hamiltonian is real, thus $C$ is plus or minus 1 and
%not the complicated expression that corresponded to the $C_{12}$ case.
As in the case of $C_{12}$ the transition of the ground state between different
spin sectors as a function of applied field can be tracked in the complex
$\lambda$ plane. However the trajectory of the zeros of
$B_{N}^{a.c.}(\lambda)$, the off diagonal element in the effective Hamiltonian
in equation (10), is different from $C_{12}$ where the zeros move from the
complex plane onto the real axis at $\lambda=1$ and $h=h_{c}$ and then move
away for $h>h_{c}$ as seen in figure 17. These values of $\lambda$ are branch
points of the energy function for the ground and first excited states. In
$C_{20}$ on the other hand close to the critical field, $h_{c}=0.514$, two
zeros of $B$ approach the real $\lambda$ axis from above and below at
$\lambda \approx 0.85$. The critical field is the value of h were the triplet
state first becomes the ground state at $\lambda=1$. As $h$ is increased
further the two zeros move along the real axis in opposite directions, as shown
in figure 26. At these values of $\lambda$ there are discontinuous jumps in the
value of $<S^{z}>$ in the ground state as opposed to the rapid variation with
$\lambda$ as seen in $C_{12}$. These jumps occur as the ground and excited
states with different values of $<S^{z}>$ switch roles when
$B_{N}^{a.c.}(\lambda)$ changes sign.

In figure 27 we plot $<S^{z}>$ for the ground state as a function of applied
magnetic field for different values of $\lambda$. The curves for $\lambda<1$
are strongly reminiscent of the results for the magnetization of the classical
approximations for $C_{20}$ and $C_{60}$ \cite{Coffey}. For $h>0.5$ there are
discontinuities in $<S^{z}>$ for different values of $\lambda$. In these
calculations starting from the field independent classical ground state the
magnetic field is part of the perturbation and is scaled by $\lambda$. So the
effective field at which discontinuities in $<S^{z}>$ take place is
$\lambda h$. In the classical approximation for $C_{20}$ there is a jump in the
magnetization as a function of $h$ at $1.43$ equal to $0.68 S_{c}$, where
$S_{c}$ is the magnitude of the classical spin. In the ground state calculated
here the jump in $<S^{z}>$ occurs for $\lambda \ge 0.72$ and grows in magnitude
until, at $\lambda=1$, the jump discontinuity in $<S^{z}>$ is $1$ for
$h_{c}=0.514$.

Different magnetic properties for the $S=\frac{1}{2}$ solutions of $C_{12}$ and
$C_{20}$ are thus seen to come from the different analytic behavior of the
wavefunctions in the complex $\lambda$ plane. Similarity of the quantum
solutions with the classical solutions suggests that this difference in
analytic structure comes from the different connectivities of the two systems.

\section{Conclusion}

We have applied perturbation theory in the strength of quantum fluctuations
around the classical ground states and found essentially exact results for the
ground and first excited states for two frustrated spin systems, $C_{12}$ and
$C_{20}$. For $C_{20}$ large orders in perturbation expansion and a high degree
of numerical precision are required to get convergence of the analytic
continuation of the matrix elements of the effective Hamiltonian and the
wavefunction coefficients calculated perturbatively.

We found that although sites in both systems are three-fold coordinated the
spectra are qualitatively different with the lowest lying excited state being a
triplet in $C_{12}$ while it is a singlet in $C_{20}$. The difference in the
behavior of the two systems is attributed to their difference in the
connectivity. The average energies per nearest neighbor bond for
$S=\frac{1}{2}$ spins, $-0.325J$ (in $C_{20}$) and $-0.3105J$ (in $C_{12}$),
taking $J_{1}=J_{2}=J$, are comparable so that the ``average'' degree of
frustration is the same for both systems. In $C_{12}$ the first excited state
arises by taking a linear combination of states each associated with a triplet
excitation of a single non-triangle bond. By comparing
$<\vec{S}_{1} \cdot \vec{S}_{4}>$ in figures 14 and 15 the non-triangle bond is
seen to take on a more strongly triplet character in the excited state. In
$C_{20}$ on the other hand there is no nearest neighbor bond which can be
singled out to describe the character of the singlet and triplet excited
states. The different values of $\vec{S}^{2}$ for states in $C_{20}$ arise from
small changes in long range correlations. In order to distinguish between
$C_{12}$ and $C_{20}$ we observe that in $C_{20}$ all bonds are members of
closed loops in which all bonds are equivalent while in $C_{12}$ the
non-triangle bonds form alternating sides with triangle bonds to make hexagons.
However characterization of connectivity by the nature of bonds forming closed
loops is not sufficient in itself since the excited state of the Heisenberg
model on a cube has a triplet excited state.

In the classical approximation it was possible to characterize the ground state
for different systems in terms of a topological number, the Skyrmion number,
and to associate this with the presence or absence of discontinuities in the
magnetization. We have shown that this difference, seen in the classical
approximation for $C_{12}$ and $C_{20}$, survives in $S^{z}$ for the exact
states of the isotropic $S=\frac{1}{2}$ Heisenberg model and that it is
associated with the analytic behavior in the complex coupling constant plane.
Although it seems clear that it is the combination of frustration and
connectivity which is responsible for the nature of the spectrum and the
response of the system to a magnetic field, how exactly these are determined is
not obvious.

The connection between frustration and the character of the excitation spectrum
is also of interest for infinite lattices. The apparent singlet nature of the
excitation spectrum in the Kagom\'e lattice arises from frustration
\cite{Mambrini,Waldtmann} and leads to a large non magnetic contribution to the
entropy of the system at low temperatures. This is an important feature in
analysis of experimental data on any physical realization of a Kagom\'e
lattice. This result for the Kagom\'e lattice may not generalize to other
frustrated lattices if the sensitivity to connectivity seen in $C_{12}$ and
$C_{20}$ is a guide to larger systems \cite{Pickett}.

The accuracy of the wavefunctions found at $\lambda=1$ suggest that the Pad\'e
approximants derived from the perturbation series are very accurate
representations of the wavefunctions and energies in the complex plane between
the origin and $\lambda=1$ and that the analytic structure of these functions
accurately represents the dependence on $\lambda$ and applied field. We have
shown how the rapid change of $<S^{z}>$ for $\lambda \simeq 1$ and
$h \simeq h_{c}$ and the discontinuity in $<S^{z}>$ for $C_{20}$ can be traced
back to the Pad\'e approximants. The ability to track this order-disorder
transition from coherent states to eigenstates of $S^{z}$ and $\vec{S}^{2}$ as
a function of $\lambda$ and applied field may be developed for richer many body
systems with the calculational resources available today.

These calculations are being extended to higher values of spin and to finite
temperatures and will be used to investigate the temperature evolution of the
properties of magnetic molecules, Mn$_{12}$, Fe$_{8}$ etc., and, in particular,
the quantum mechanical description of the large moments presently used in their
analysis \cite{DeRaedt}.

D. Coffey thanks G. Baker, R. Singh and S. Trugman for discussions on some of
the points raised in the calculations and D. C. Mattis for comments and
suggestions on the manuscript. The authors thank N. Bock and M. Jones for
technical assistance. Numerical calculations were performed on the machines of
the Center of Computational Research at SUNY Buffalo.

%
%references
%

%
%table
%

%\newpage
%\begin{table}
%\caption{$S^{z}$ as a function of the magnetic field h for $C_{12}$ for
%         $\lambda = 1$ - degenerate case}
%\begin{tabular}{ccc}
%magnetic field h&double precision&quadruple precision\\
%\hline
%%0.5&0.010413161057868&-0.0000049170645882549309828063558589\\0.7&0.795069772152895&0.9999975452285270405533531862872460\\0.8&0.987309248485774&0.9999932839979933632835077152044482\\1.0&0.874918765455824&1.2769104825079011853982581284645494\\
%0.500&0.0104132&1.711482e-05\\0.600&-&1.525703e-03\\0.685&-&1.037872e-02\\0.690&-&0.970537\\0.700&0.795070&0.999760\\
%\end{tabular}
%\label{4/22/98}
%\end{table}

%\newpage
%\begin{table}
%\caption{$S^{z}$ as a function of the magnetic field h for $C_{12}$ for
%         $\lambda = 1$ - degenerate case}
%\begin{tabular}{cc}
%magnetic field h&$S^{z}$\\
%\hline
%0.688&0.99999985725167095\\0.690&0.99999999975072296\\
%\end{tabular}
%\label{9/29/99}
%\end{table}

%\begin{table}
%\caption{$S^{z}$ as a function of the magnetic field h for $C_{12}$ for
%         $\lambda = 1$ - non degenerate case}
%\begin{tabular}{ccc}
%magnetic field h&double precision&quadruple precision\\
%\hline
%0.2&0.204580597530098&-0.2948770749515670332794288155619208\\0.5&-0.004086174003988&-\\0.7&0.819652751766531&0.9890377259226577671775657235557308\\0.8&0.987486195657912&0.9999399646633292468306555518595141\\
%\end{tabular}
%\label{4/22/98}
%\end{table}

%
%figures
%

\newpage
%\begin{figure}
%\begin{center}
%\epsfig{file=energyconv.eps,width=3in}
%\end{center}
%\caption{11/22/98 - Convergence rate for the ground state energy of $C_{12}$
%         for $\alpha=1$ ( double precision )}
%\label{11/22/98}
%\end{figure}

\begin{figure}
\begin{center}
\epsfig{file=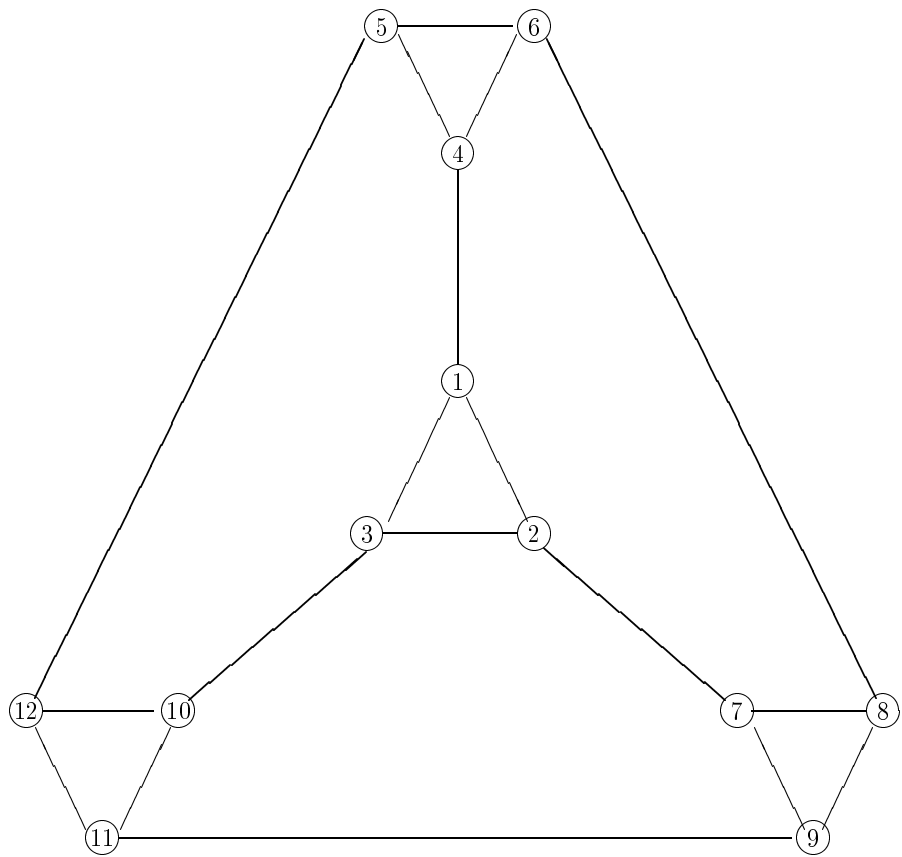,width=6.8in}
\end{center}
\caption{Space configuration of $C_{12}$. Intra-triangle bonds are called
         $J_{1}$, while inter-triangle bonds are called $J_{2}$.}
\label{}
\end{figure}

\begin{figure}
\begin{center}
\epsfig{file=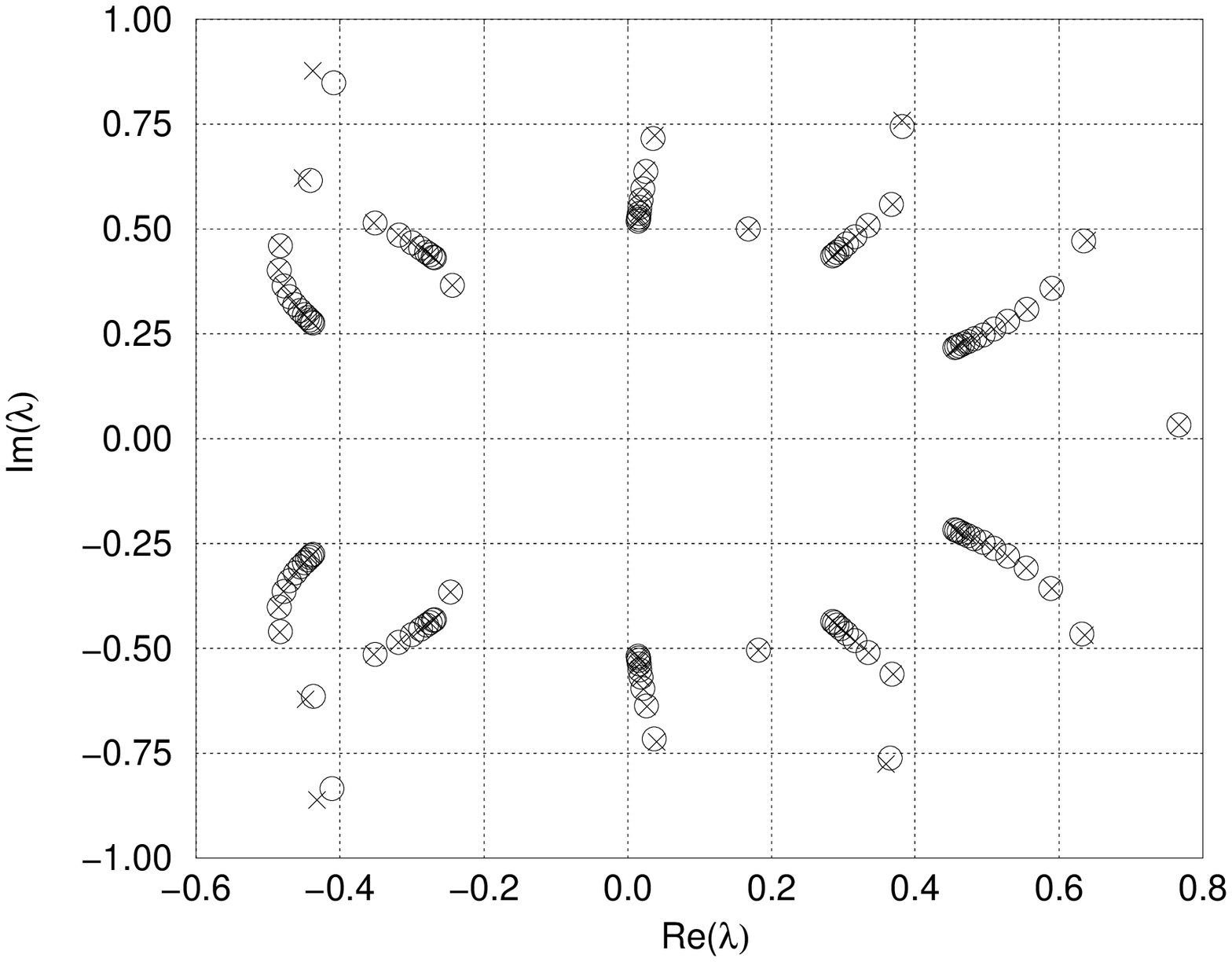,width=4.2in}
\end{center}
\caption{Zeros and poles of the Pad\'e approximant of the diagonal element
         $A_{N}$ of the effective Hamiltonian for $\alpha=1$ in $C_{12}$ with
         220 orders of perturbation theory used, $\circ$ : zeros , $\times$ :
         poles.}
\label{9/12/00}
\end{figure}

\begin{figure}
\begin{center}
\epsfig{file=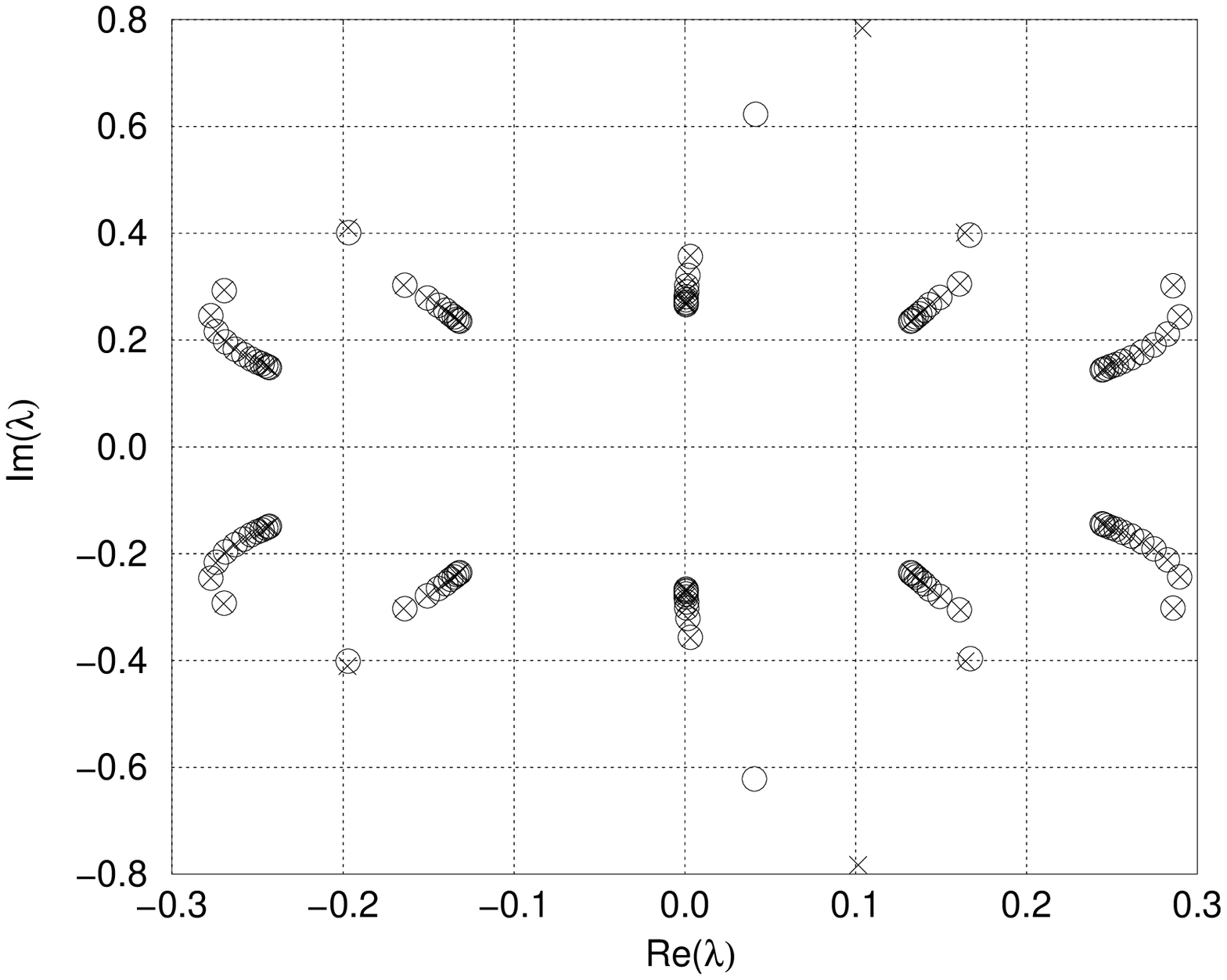,width=4.2in}
\end{center}
\caption{Zeros and poles of the Pad\'e approximant of the diagonal element
         $A_{N}$ of the effective Hamiltonian for $\alpha=2$ in $C_{12}$ with
         220 orders of perturbation theory used, $\circ$ : zeros , $\times$ :
         poles.}
\label{9/12/00}
\end{figure}

\begin{figure}
\begin{center}
\epsfig{file=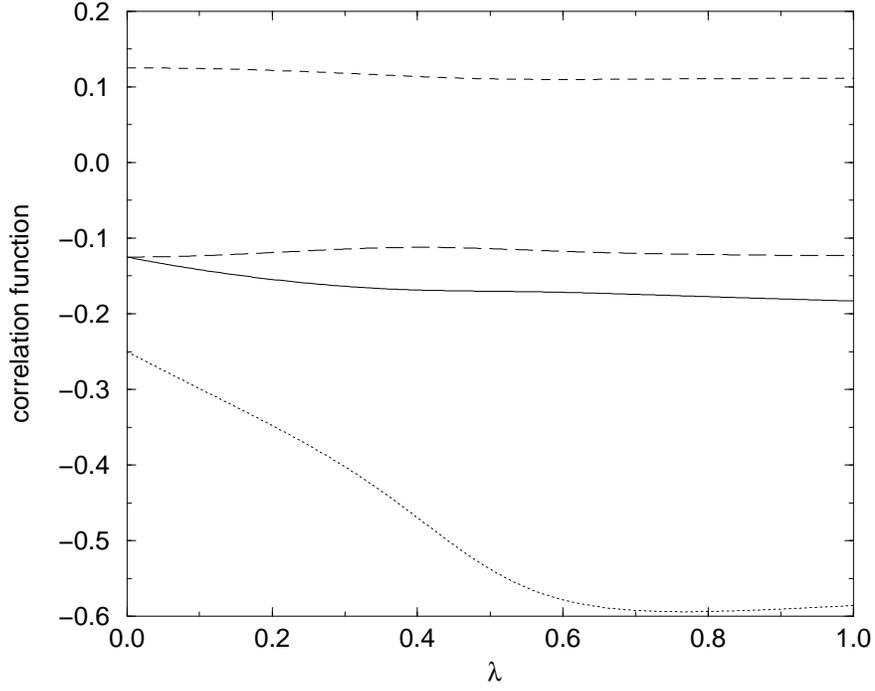,width=3.8in}
\end{center}
\vspace{20pt}
\caption{Correlation functions for $J_{2}=1$ in the ground state of $C_{12}$ at
         $h=0$ : solid line : $<\vec{S}_{1}\cdot\vec{S}_{2}>$ , dotted line :
         $<\vec{S}_{1}\cdot\vec{S}_{4}>$, dashed line :
         $<\vec{S}_{1}\cdot\vec{S}_{5}>$ , long dashed line :
         $<\vec{S}_{1}\cdot\vec{S}_{8}>$}
\label{7/12/00}
\end{figure}

\begin{figure}
\begin{center}
\epsfig{file=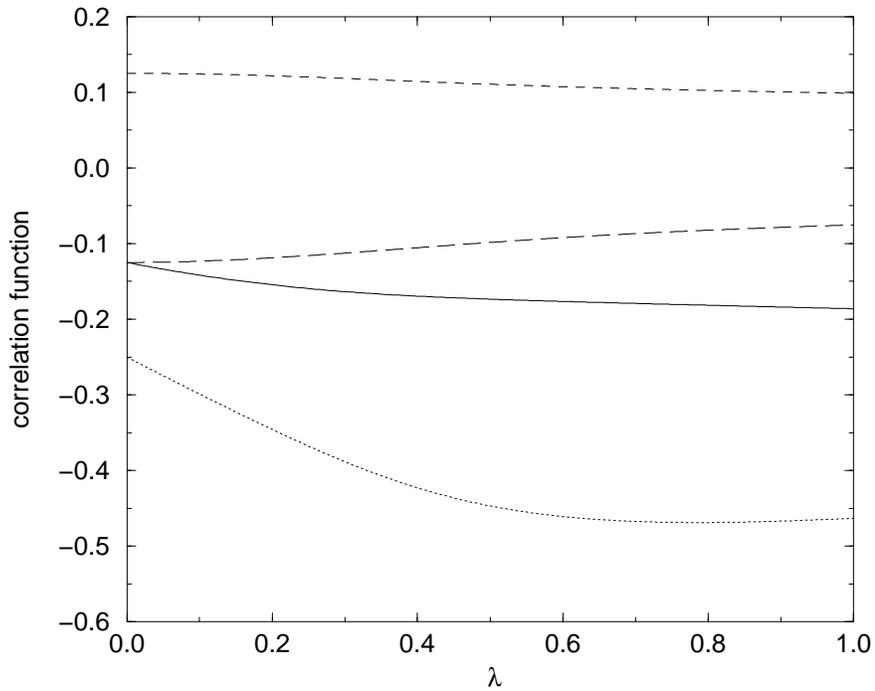,width=3.8in}
\end{center}
\vspace{20pt}
\caption{Correlation functions for $J_{2}=1$ in the first excited state of
         $C_{12}$ at $h=0$ : solid line : $<\vec{S}_{1}\cdot\vec{S}_{2}>$ ,
         dotted line : $<\vec{S}_{1}\cdot\vec{S}_{4}>$, dashed line :
         $<\vec{S}_{1}\cdot\vec{S}_{5}>$ , long dashed line :
         $<\vec{S}_{1}\cdot\vec{S}_{8}>$}
\label{7/12/00}
\end{figure}

\begin{figure}
\begin{center}
\epsfig{file=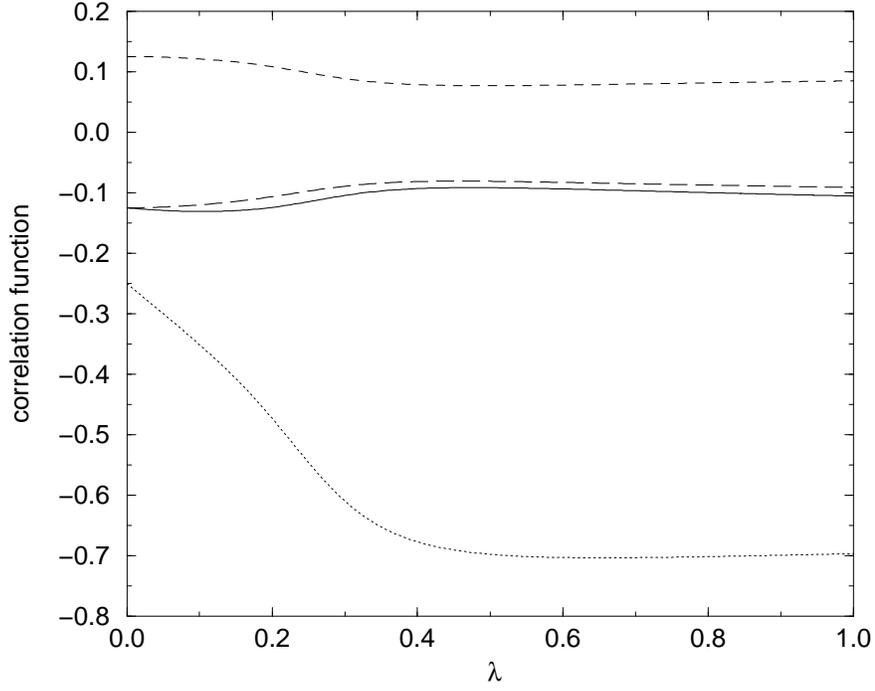,width=3.8in}
\end{center}
\vspace{20pt}
\caption{Nearest neighbor correlation functions for $J_{2}=2$ in the ground
         state of $C_{12}$ at $h=0$ : solid line :
         $<\vec{S}_{1}\cdot\vec{S}_{2}>$ , dotted line :
         $<\vec{S}_{1}\cdot\vec{S}_{4}>$, dashed line :
         $<\vec{S}_{1}\cdot\vec{S}_{5}>$ , long dashed line :
         $<\vec{S}_{1}\cdot\vec{S}_{8}>$}
\label{7/13/00}
\end{figure}

\begin{figure}
\begin{center}
\epsfig{file=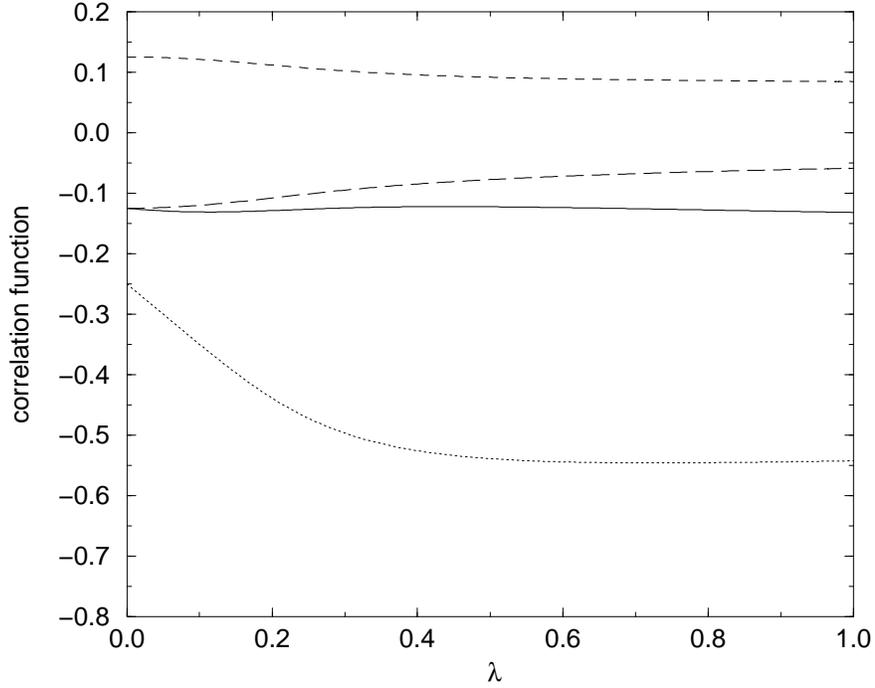,width=3.8in}
\end{center}
\vspace{20pt}
\caption{Nearest neighbor correlation functions for $J_{2}=2$ in the $S=1$ ,
         $S^{z}=0$ first excited state of $C_{12}$ at $h=0$ : solid line :
         $<\vec{S}_{1}\cdot\vec{S}_{2}>$ , dotted line :
         $<\vec{S}_{1}\cdot\vec{S}_{4}>$, dashed line :
         $<\vec{S}_{1}\cdot\vec{S}_{5}>$ , long dashed line :
         $<\vec{S}_{1}\cdot\vec{S}_{8}>$}
\label{7/13/00}
\end{figure}

%\begin{figure}
%\begin{center}
%\epsfig{file=S1S2.eps,width=3in}
%\end{center}
%\caption{9/24/98 - Nearest neighbor correlation function $S_{1}S_{2}$ for
%         $C_{12}$ for $J_{1}=J_{2}$}
%\label{9/24/98}
%\end{figure}

%\begin{figure}
%\begin{center}
%\epsfig{file=S1S4.eps,width=3in}
%\end{center}
%\caption{9/24/98 - Nearest neighbor correlation function $S_{1}S_{4}$ for
%         $C_{12}$ for $J_{1}=J_{2}$}
%\label{9/24/98}
%\end{figure}

%\begin{figure}
%\begin{center}
%\epsfig{file=S1S5.eps,width=3in}
%\end{center}
%\caption{9/24/98 - Next nearest neighbor correlation function $S_{1}S_{5}$ for
%         $C_{12}$ for $J_{1}=J_{2}$}
%\label{9/24/98}
%\end{figure}

%\begin{figure}
%\begin{center}
%\epsfig{file=S1S8.eps,width=3in}
%\end{center}
%\caption{9/24/98 - Correlation function $S_{1}S_{8}$ for $C_{12}$ for
%         $J_{1}=J_{2}$}
%\label{9/24/98}
%\end{figure}

\begin{figure}
\begin{center}
\epsfig{file=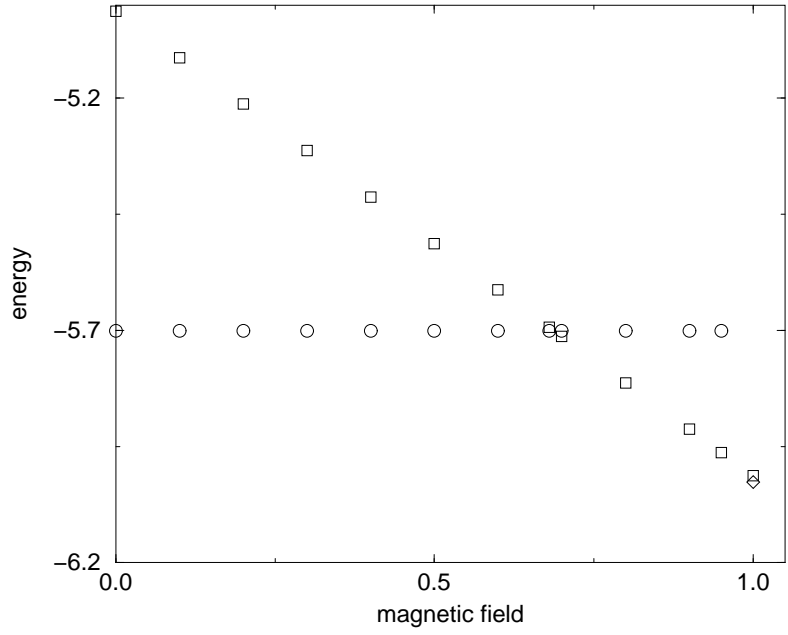,width=4.1in}
\end{center}
\vspace{20pt}
\caption{Ground state and first excited state energy as a function of applied
         magnetic field for $C_{12}$, $\circ$ : $S^{z}=0$, $\Box$ : $S^{z}=1$,
         $\diamond$ : $S^{z}=2$.}
\label{8/3/00}
\end{figure}

\begin{figure}
\begin{center}
\epsfig{file=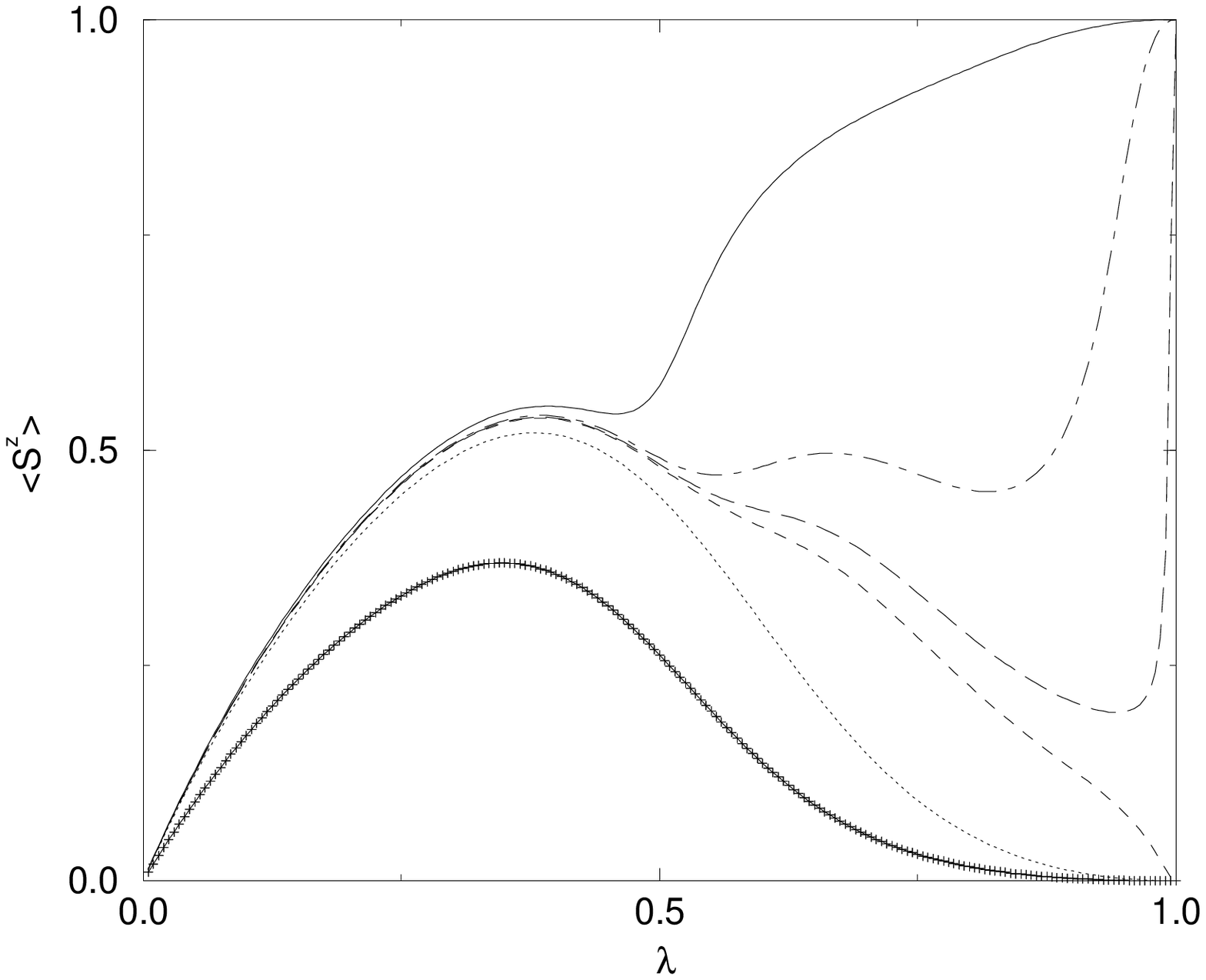,width=3.8in}
\end{center}
\vspace{20pt}
\caption{$S^{z}$ vs. $\lambda$ for $C_{12}$ for a magnetic field h ( ground
         state ) starting from the magnetic field independent classical ground
         state , $J_{2}=1.0$ , solid-plus line : h=0.500 , dotted line :
         h=0.670 , dashed line : h=0.687 , long dashed line : h=0.688 ,
         dot-dashed line : h=0.690 , solid line : h=0.700.}
\label{3/10/00}
\end{figure}

\begin{figure}
\begin{center}
\epsfig{file=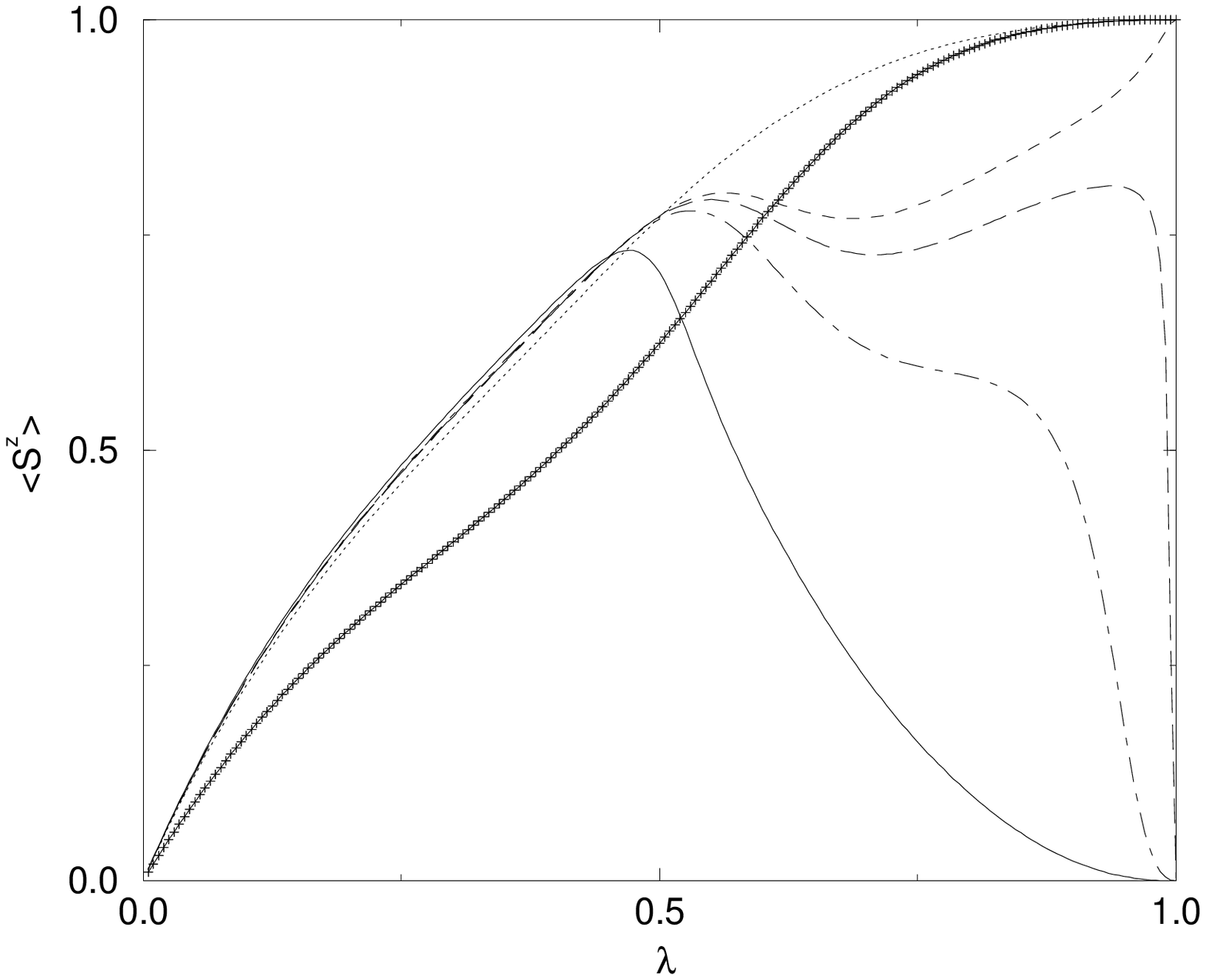,width=3.7in}
\end{center}
\vspace{20pt}
\caption{$S^{z}$ vs. $\lambda$ for $C_{12}$ for a magnetic field h ( excited
         state ) starting from the magnetic field independent classical ground
         state , $J_{2}=1.0$ , solid-plus line : h=0.500 , dotted line :
         h=0.670 , dashed line : h=0.687 , long dashed line : h=0.688 ,
         dot-dashed line : h=0.690 , solid line : h=0.700.}
\label{3/10/00}
\end{figure}

\begin{figure}
\begin{center}
\epsfig{file=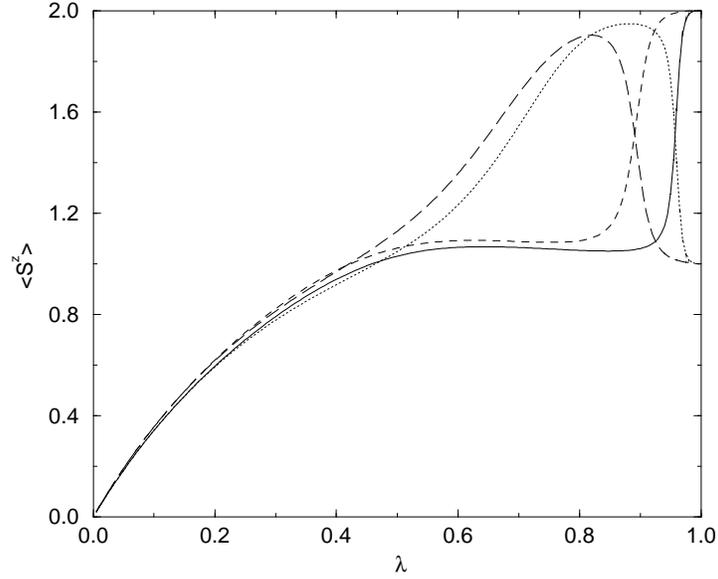,width=3.7in}
\end{center}
\vspace{20pt}
\caption{$S^{z}$ vs. $\lambda$ for $C_{12}$ for a magnetic field h
         starting from the magnetic field independent classical ground state ,
         $J_{2}=1.0$ , solid line : h=1.010 , ground state , dotted line :
         h=1.010 , excited state , dashed line : h=1.050 , ground state , long
         dashed line : h=1.050 , excited state.}
\label{8/14/00}
\end{figure}

\begin{figure}
\begin{center}
\epsfig{file=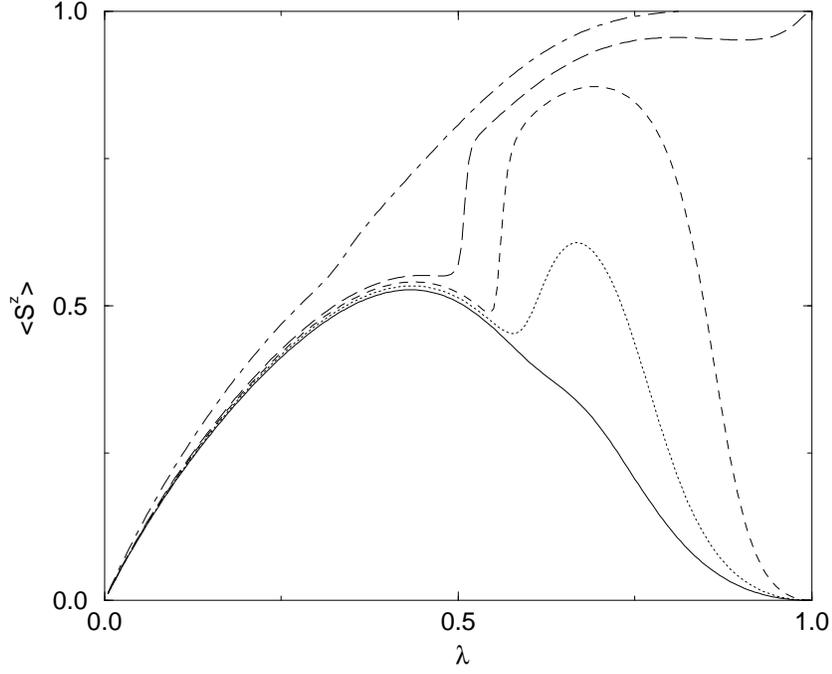,width=3.7in}
\end{center}
\vspace{20pt}
\caption{$S^{z}$ vs. $\lambda$ for $C_{12}$ for a magnetic field h starting
         from the magnetic field independent classical ground state ,
         $J_{2}=0.8$ , solid line : h=0.530 , dotted line : h=0.535 , dashed
         line : h=0.540 , long dashed line : h=0.547 , dot-dashed line :
         h=0.600.}
\label{3/12/00}
\end{figure}

\begin{figure}
\begin{center}
\epsfig{file=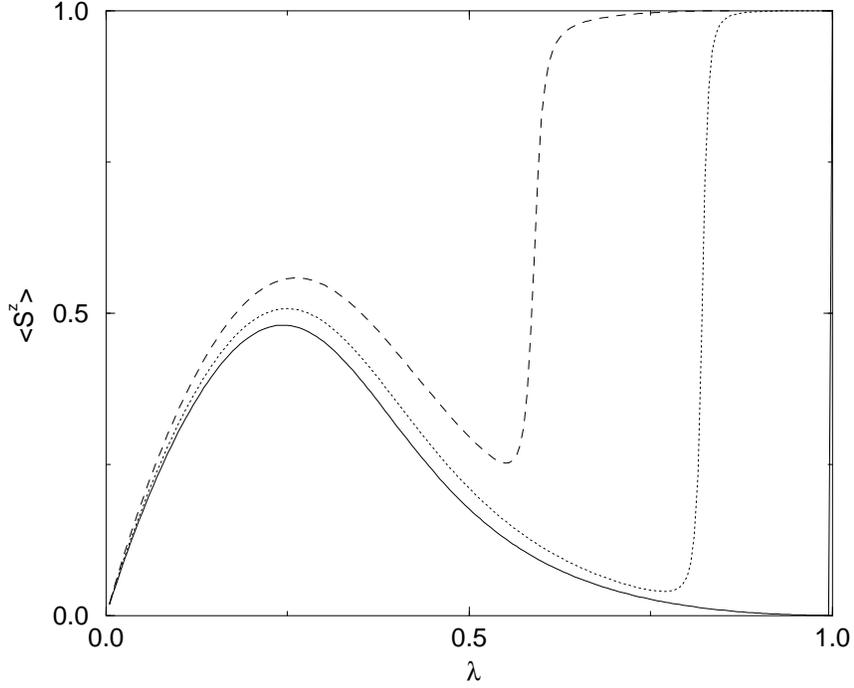,width=3.8in}
\end{center}
\vspace{20pt}
\caption{$S^{z}$ vs. $\lambda$ for $C_{12}$ for a magnetic field h starting
         from the magnetic field independent classical ground state ,
         $J_{2}=1.8$ , solid line : h=1.3423 , dotted line : h=1.400 , dashed
         line : h=1.500.}
\label{3/12/00}
\end{figure}

\begin{figure}
\begin{center}
\epsfig{file=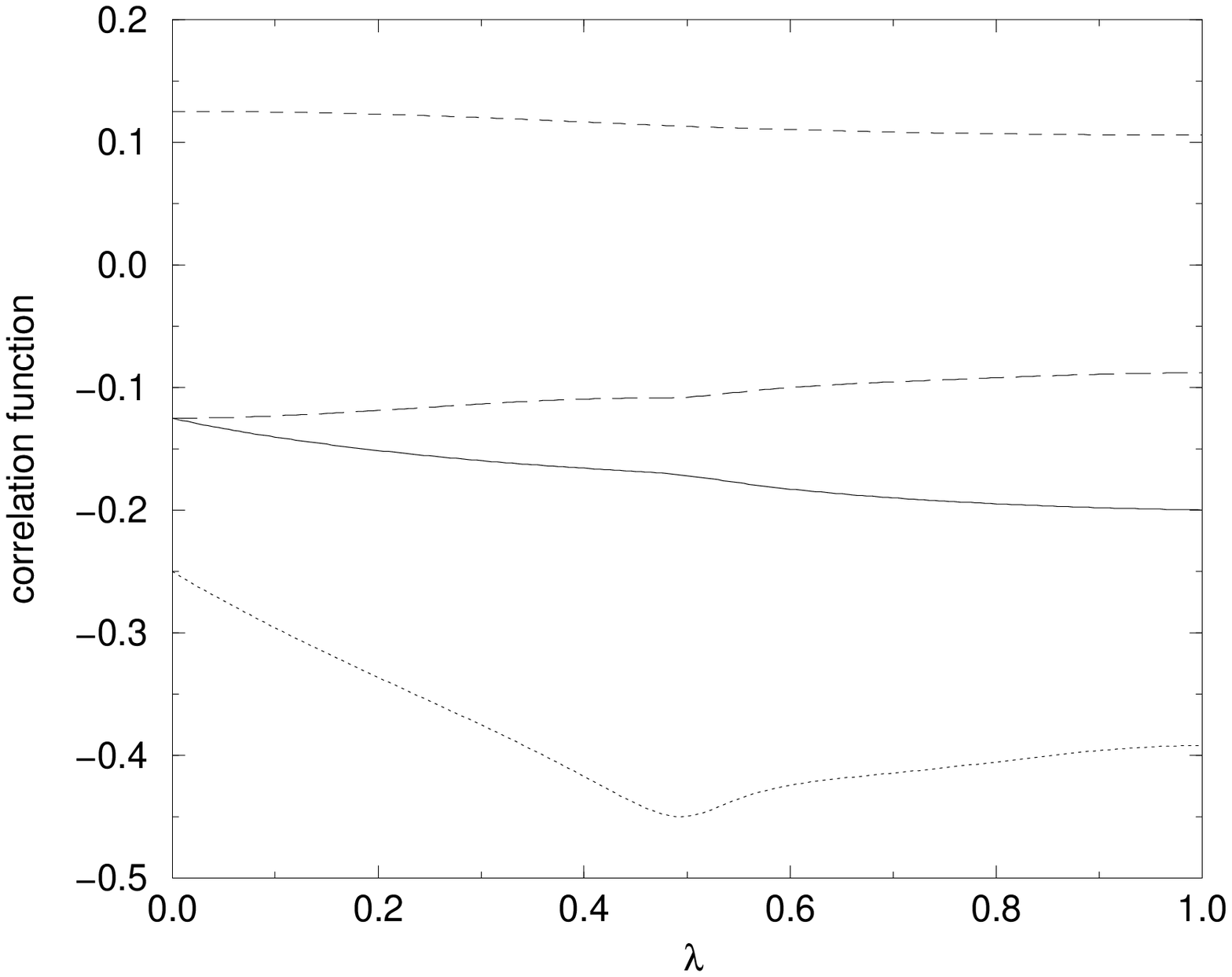,width=3.8in}
\end{center}
\vspace{20pt}
\caption{Nearest neighbor correlation functions for $C_{12}$ for $J_{2}=1$ ,
         $h=0.7$ , ground state : solid line :
         $<\vec{S}_{1}\cdot\vec{S}_{2}>$ , dotted line :
         $<\vec{S}_{1}\cdot\vec{S}_{4}>$, dashed line :
         $<\vec{S}_{1}\cdot\vec{S}_{5}>$ , long dashed line :
         $<\vec{S}_{1}\cdot\vec{S}_{8}>$}
\label{7/18/00}
\end{figure}

\begin{figure}
\begin{center}
\epsfig{file=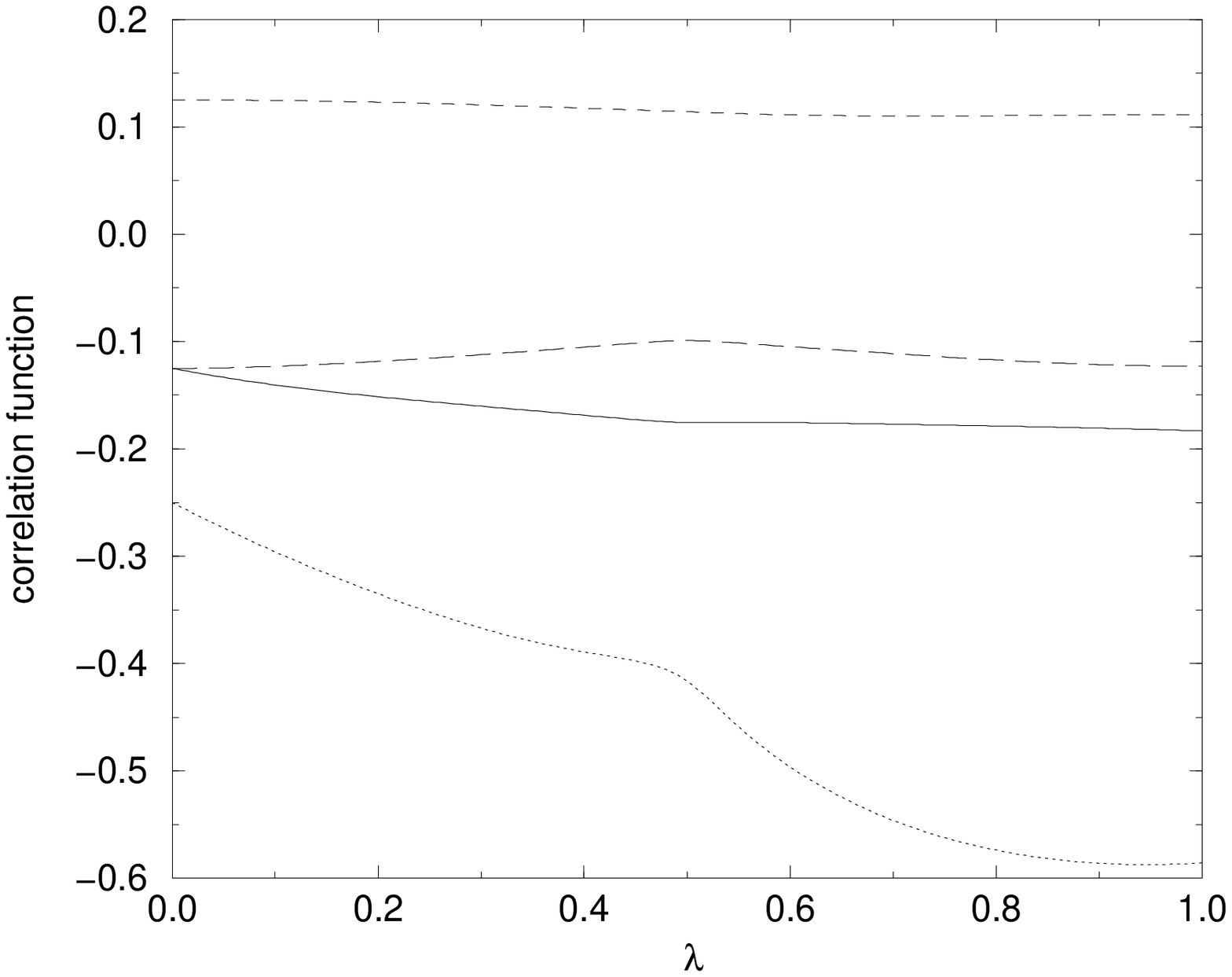,width=3.8in}
\end{center}
\vspace{20pt}
\caption{Nearest neighbor correlation functions for $C_{12}$ for $J_{2}=1$ ,
         $h=0.7$ , excited state : solid line :
         $<\vec{S}_{1}\cdot\vec{S}_{2}>$ , dotted line :
         $<\vec{S}_{1}\cdot\vec{S}_{4}>$, dashed line :
         $<\vec{S}_{1}\cdot\vec{S}_{5}>$ , long dashed line :
         $<\vec{S}_{1}\cdot\vec{S}_{8}>$}
\label{7/18/00}
\end{figure}

\begin{figure}
\begin{center}
\epsfig{file=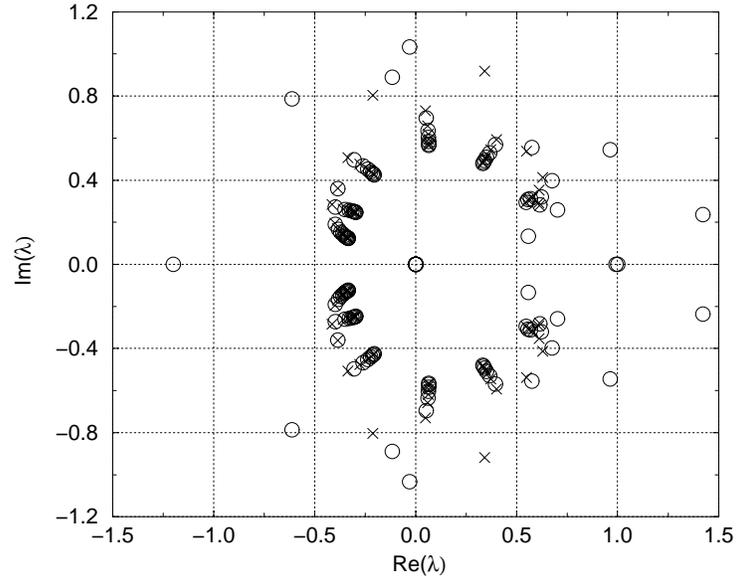,width=3.8in}
\end{center}
\vspace{20pt}
\caption{Zeros and poles of the Pad\'e approximant of the square root in
         $C_{12}$ with 220 orders of perturbation theory used, magnetic field
         h=0.688, $\circ$ : zeros , $\times$ : poles.}
\label{3/13/00}
\end{figure}

\begin{figure}
\begin{center}
\epsfig{file=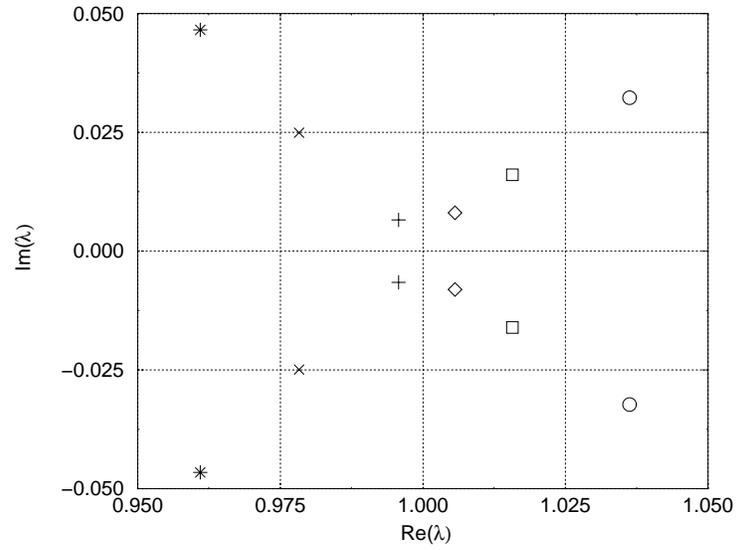,width=3.8in}
\end{center}
\vspace{20pt}
\caption{Zeros of the Pad\'e approximant for the square root using 270 orders
         of perturbation theory for various magnetic fields , $\circ$ :
         h=0.686 , $\Box$ : h=0.687 , $\diamond$ : h=0.6875 , + : h=0.688 ,
         $\times$ : h=0.689 , $\ast$ : h=0.690.}
\label{9/13/00}
\end{figure}

\begin{figure}
\begin{center}
\epsfig{file=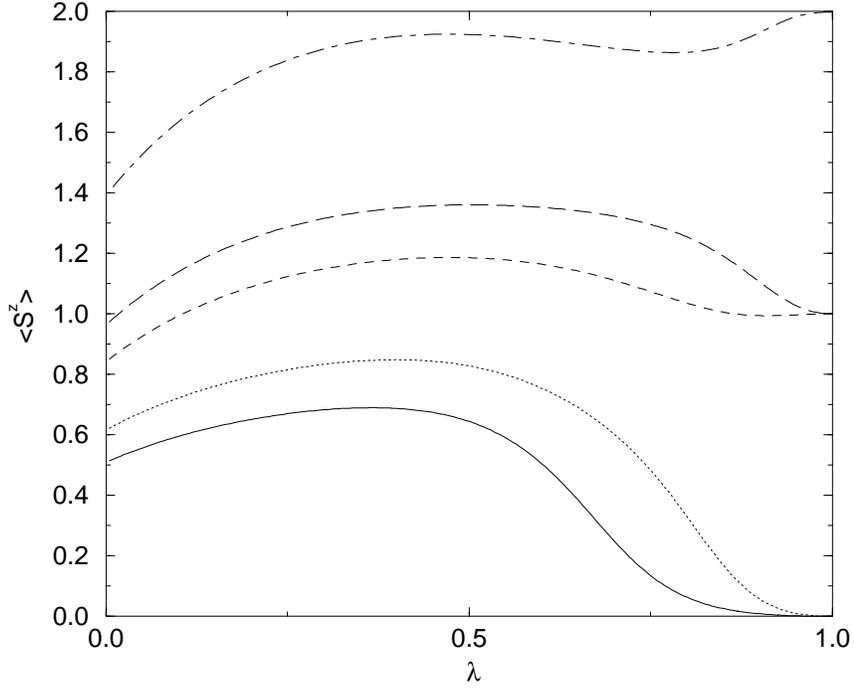,width=3.8in}
\end{center}
\vspace{20pt}
\caption{$S^{z}$ vs. $\lambda$ for $C_{12}$ for a magnetic field h starting
         from the magnetic field dependent classical ground state ,
         $J_{2}=1.0$ , solid line : h=0.5 , dotted line : h=0.6 , dashed line :
         h=0.8 , long dashed line : h=0.9 , dot-dashed line : h=1.200.}
\label{3/12/00}
\end{figure}

\newpage

\begin{figure}
\begin{center}
\epsfig{file=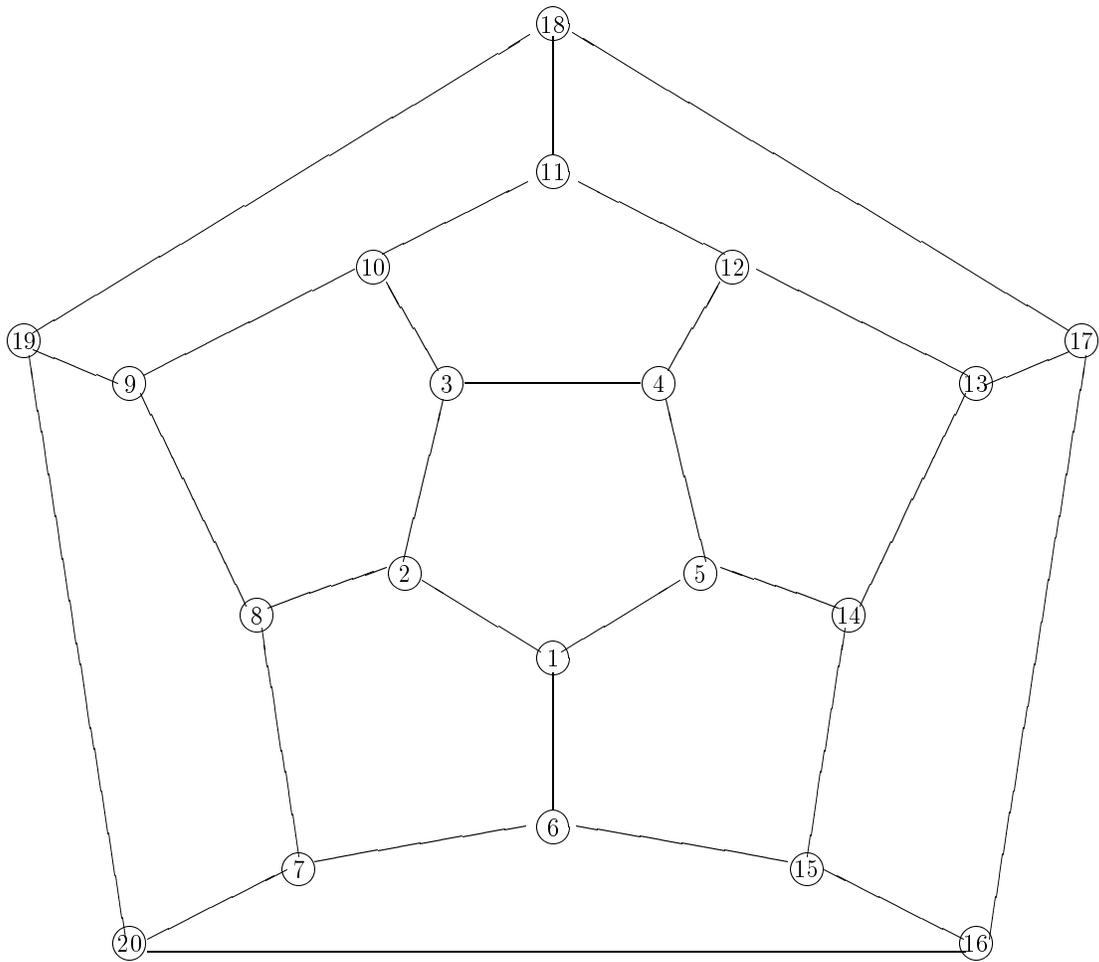,width=6.8in}
\end{center}
\caption{Space configuration of $C_{20}$. All the bonds are equal to $J$.}
\label{}
\end{figure}

\begin{figure}
\begin{center}
\epsfig{file=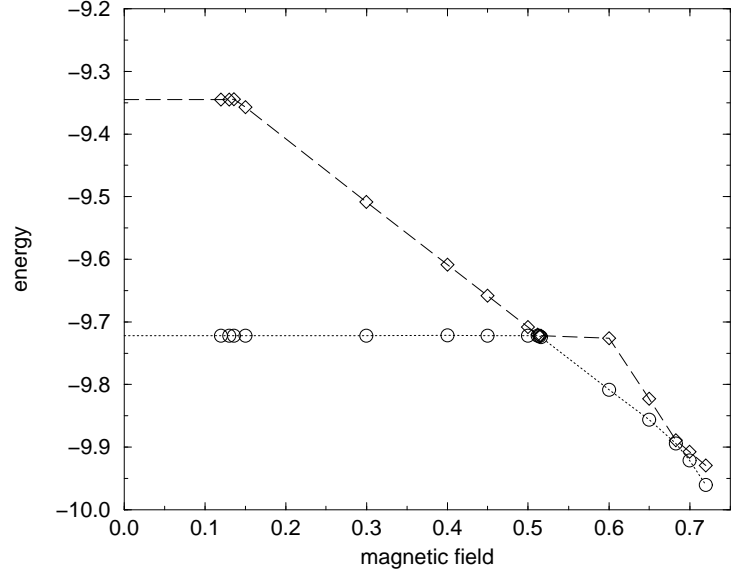,width=3.8in}
\end{center}
\vspace{20pt}
\caption{Ground state and first excited state energy for different values of
         applied magnetic field for $C_{20}$ , $\circ$ : ground state ,
         $\diamond$ : excited state. The dotted line is a fit for the ground
         state, and the long dashed line for the excited state.}
\label{8/17/00}
\end{figure}

\begin{figure}
\begin{center}
\epsfig{file=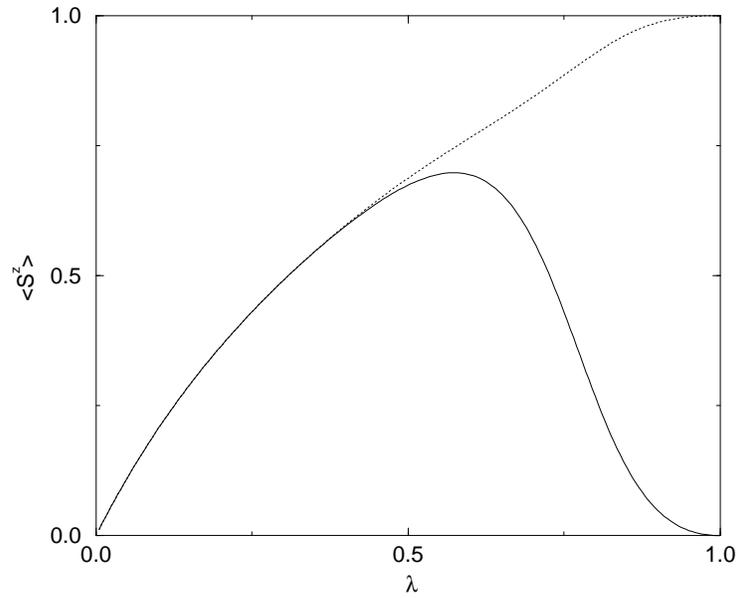,width=3.8in}
\end{center}
\vspace{20pt}
\caption{$<S^{z}>$ as a function of $\lambda$ for the ground state and first
         excited state of $C_{20}$ for a magnetic field h = 0.4 starting from
         the magnetic field independent classical ground state , solid line :
         ground state , dotted line : excited state.}
\label{6/7/00}
\end{figure}

\begin{figure}
\begin{center}
\epsfig{file=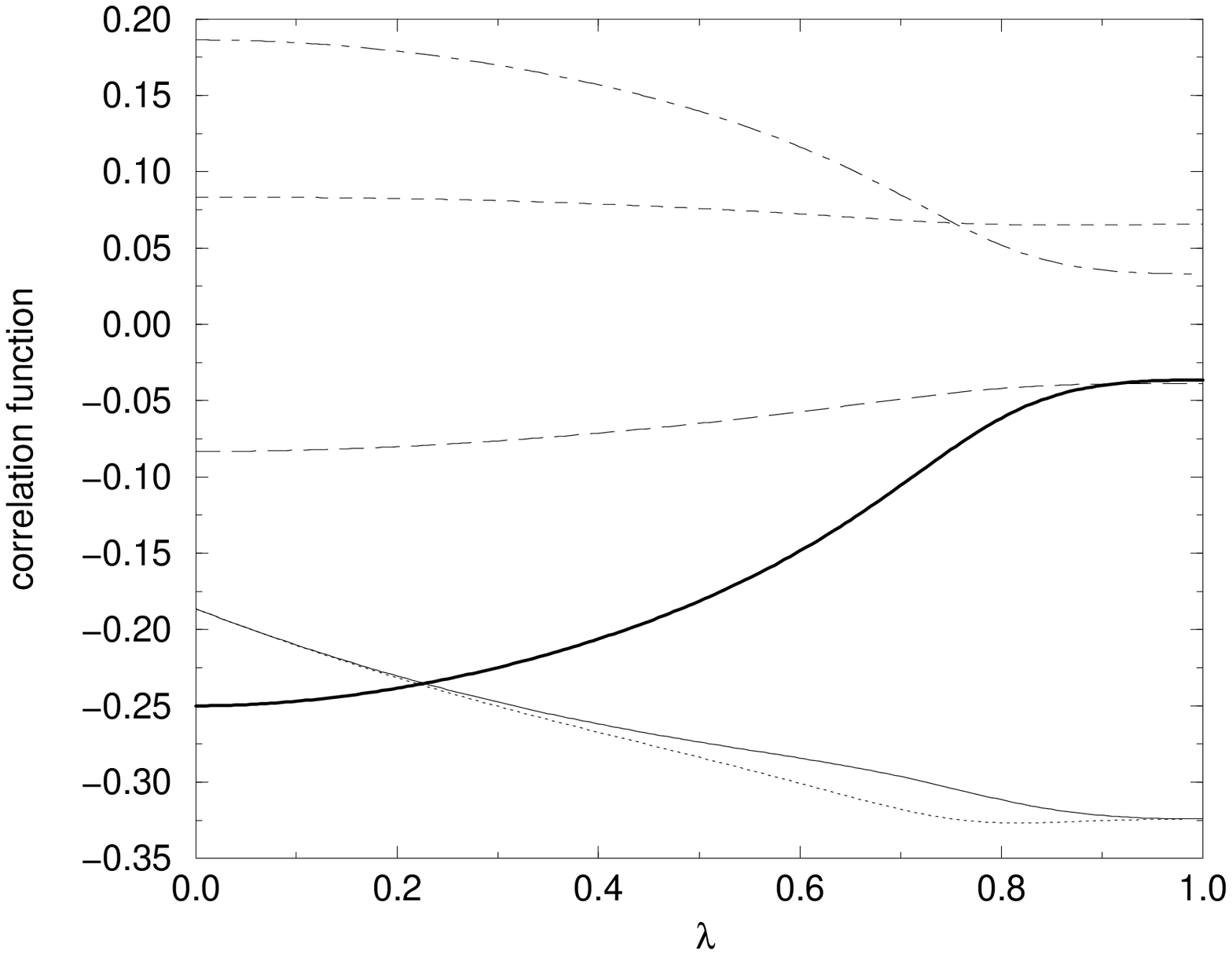,width=3.8in}
\end{center}
\vspace{20pt}
\caption{Spin correlations in the ground state of $C_{20}$ as a function of
         $\lambda$ for h=0.4: solid line : $<\vec S_{1} \cdot \vec S_{2}>$,
         dotted line : $<\vec S_{1} \cdot \vec S_{6}>$, dashed line :
         $<\vec S_{1} \cdot \vec S_{3}>$, long dashed line :
         $<\vec S_{1} \cdot \vec S_{9}>$, dot-dashed line:
         $<\vec S_{1} \cdot \vec S_{11}>$, thick solid line:
         $<\vec S_{1} \cdot \vec S_{18}>$.}
\label{8/10/00}
\end{figure}

\begin{figure}
\begin{center}
\epsfig{file=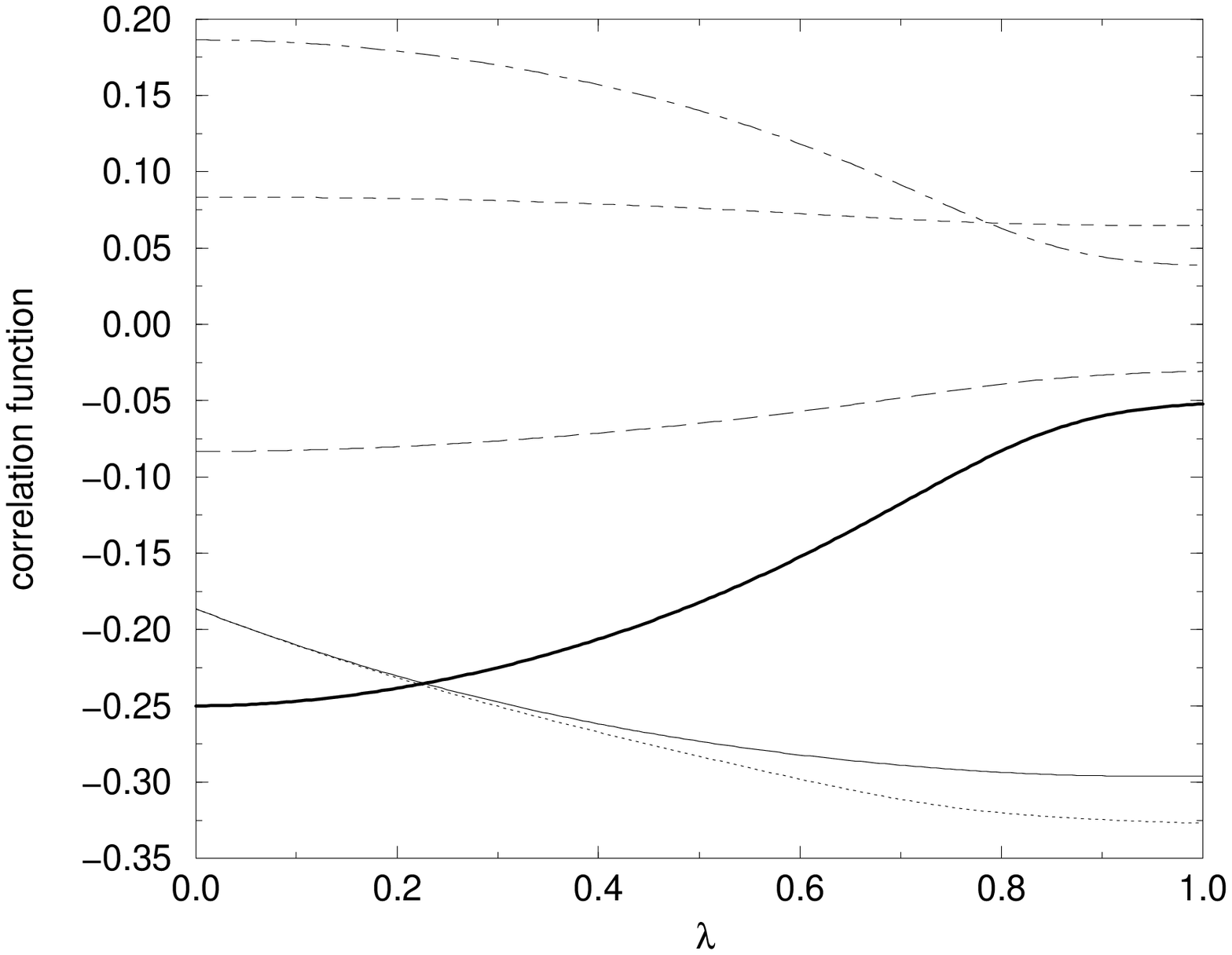,width=3.8in}
\end{center}
\vspace{20pt}
\caption{Spin correlations in the first excited state of $C_{20}$ as a function
         of $\lambda$ for h=0.4: solid line : $<\vec S_{1} \cdot \vec S_{2}>$,
         dotted line : $<\vec S_{1} \cdot \vec S_{6}>$, dashed line :
         $<\vec S_{1} \cdot \vec S_{3}>$, long dashed line :
         $<\vec S_{1} \cdot \vec S_{9}>$, dot-dashed line:
         $<\vec S_{1} \cdot \vec S_{11}>$, thick solid line:
         $<\vec S_{1} \cdot \vec S_{18}>$.}
\label{8/10/00}
\end{figure}

\begin{figure}
\begin{center}
\epsfig{file=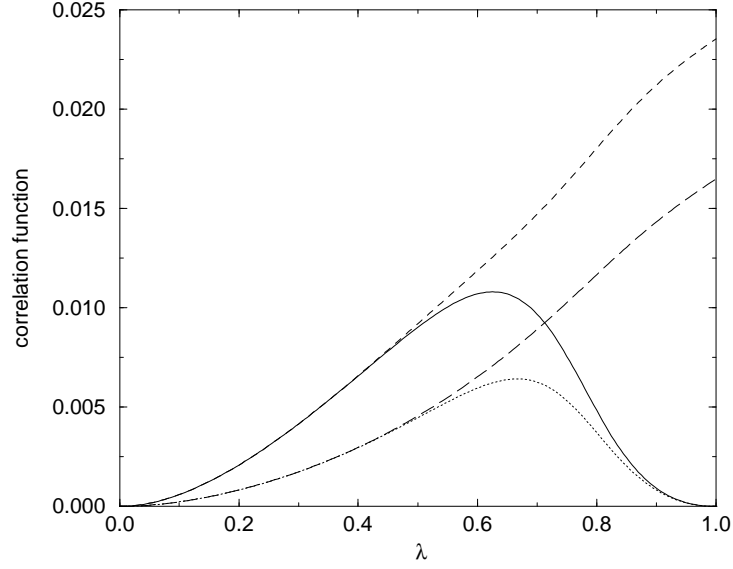,width=3.8in}
\end{center}
\vspace{20pt}
\caption{Ground and excited state spin correlations for $h=0.4$ as a function
         of $\lambda$ for $C_{20}$ at sites i=1 and i=6: solid line :
         $<\vec S_{1} \cdot \vec h>$ in the ground state, dotted line :
         $<\vec S_{6} \cdot \vec h>$ in the ground state, dashed line :
         $<\vec S_{1} \cdot \vec h>$ in the excited state, long dashed line :
         $<\vec S_{6} \cdot \vec h>$ in the excited state.}
\label{8/10/00}
\end{figure}

\begin{figure}
\begin{center}
\epsfig{file=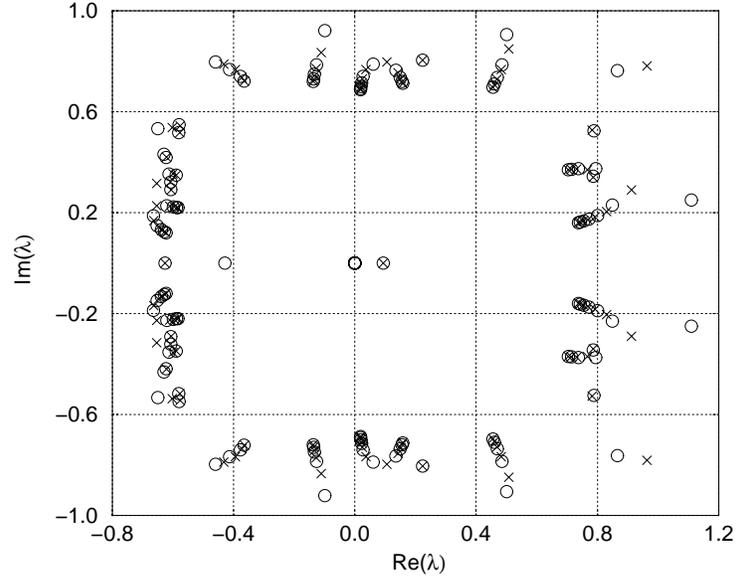,width=3.8in}
\end{center}
\vspace{20pt}
\caption{Zeros ($\circ$) and poles ($\times$) of the Pad\'e approximant
         $B_{N}^{a.c.}(\lambda)$ in the complex $\lambda$ plane for $C_{20}$ in
         an applied field $h=0.3$, 240 orders of perturbation theory were
         used.}
\label{8/3/00}
\end{figure}

\begin{figure}
\begin{center}
\epsfig{file=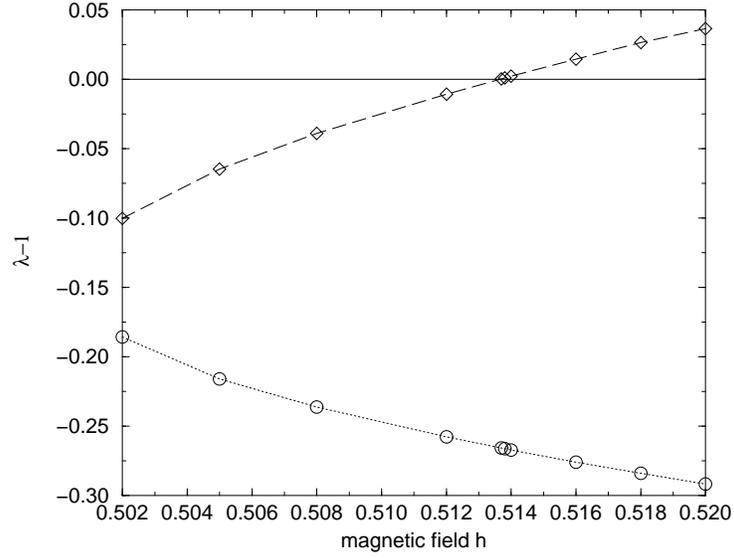,width=3.8in}
\end{center}
\vspace{20pt}
\caption{Zeros of the Pad\'e approximant $B_{N}^{a.c.}(\lambda)$ along the
         $\lambda$ real axis for $C_{20}$. 240 orders of perturbation theory
         were used and $\circ$ and $\diamond$ are the values calculated. The
         dotted and long-dashed lines are fits.}
%for magnetic fields h=0.5 ($\circ$), h=0.502
%                 ($\Box$), h=0.505 ($\diamond$), h=0.508 ($\times$), h=0.512
%                 ($\ast$), h=0.514 ($+$), h=0.516 ($\triangle$).}
\label{9/13/00}
\end{figure}

\begin{figure}
\begin{center}
\epsfig{file=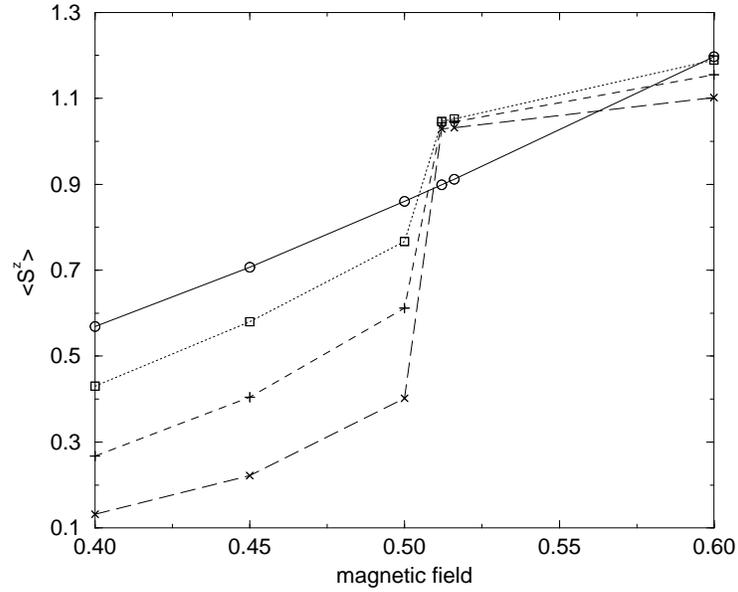,width=3.8in}
\end{center}
\vspace{20pt}
\caption{$<S^{z}>$ as a function of magnetic field for the ground state of
         $C_{20}$ for various $\lambda$'s : $\circ$ : $\lambda=0.70$ , $\Box$ :
         $\lambda=0.75$ , + : $\lambda=0.80$ , $\times$ : $\lambda=0.85$. The
         lines are fits.}
\label{10/9/00}
\end{figure}

%\begin{figure}
%\begin{center}
%\epsfig{file=energyC20.eps,width=3in}
%\end{center}
%\caption{11/22/98 - Ground state energy as a function of the perturbation
%         parameter $\lambda$ for $C_{20}$}
%\label{11/22/98}
%\end{figure}

%\begin{figure}
%\begin{center}
%\epsfig{file=energyconvC20.eps,width=3in}
%\end{center}
%\caption{11/22/98 - Convergence rate for the ground state energy of $C_{20}$
%                    ( double precision )}
%\label{11/22/98}
%\end{figure}

%\begin{figure}
%\begin{center}
%\epsfig{file=rp.eps,width=3in}
%\end{center}
%\caption{11/22/98 - Roots and poles of $A_{N}$ for $C_{20}$, $\Diamond$ :
%         roots , + : poles.}
%\label{11/22/98}
%\end{figure}

%\begin{figure}
%\begin{center}
%\epsfig{file=energyC20hfieldld.eps,width=3in}
%\end{center}
%\caption{4/14/99 - Ground state energy vs. $\lambda$ for $C_{20}$ for a
%                   magnetic field h, $\Diamond$ : h=0.5 , + : h=0.7 , $\Box$ :
%                   h=0.8 , $\times$ : h=1.0 , $\triangle$ : h=1.43}
%\label{4/14/99}
%\end{figure}

\end{document}